%% file: main.tex
\documentclass[aps,prc,longbibliography,nofootinbib,showkeys,superscriptaddress,twocolumn,longbibliography,floatfix]{revtex4-2}

\usepackage{hyperref}
\usepackage{graphicx} %

\usepackage{xspace}
\usepackage{comment}
\usepackage{multirow}
\usepackage{array,booktabs}

\usepackage{amsmath,amssymb}

\usepackage{rotating}
\usepackage{bm}

\usepackage[normalem]{ulem}

\graphicspath{ {./figures/} }

\input{commands}

\begin{document}

\title{Bayesian Inference analysis of jet quenching using inclusive jet and hadron suppression measurements}

\author{R.~Ehlers}
\affiliation{Department of Physics, University of California, Berkeley CA 94270.}
\affiliation{Nuclear Science Division, Lawrence Berkeley National Laboratory, Berkeley CA 94270.}

\author{Y.~Chen}
\affiliation{Laboratory for Nuclear Science, Massachusetts Institute of Technology, Cambridge MA 02139.}
\affiliation{Department of Physics, Massachusetts Institute of Technology, Cambridge MA 02139.}
\affiliation{Department of Physics and Astronomy, Vanderbilt University, Nashville TN 37235.}

\author{J.~Mulligan}
\affiliation{Department of Physics, University of California, Berkeley CA 94270.}
\affiliation{Nuclear Science Division, Lawrence Berkeley National Laboratory, Berkeley CA 94270.}

\author{Y.~Ji}
\affiliation{Department of Statistical Science, Duke University, Durham NC 27708.}

\author{A.~Kumar}
\affiliation{Department of Physics, University of Regina, Regina, SK S4S 0A2, Canada.}
\affiliation{Department of Physics, McGill University, Montr\'{e}al QC H3A\,2T8, Canada.}
\affiliation{Department of Physics and Astronomy, Wayne State University, Detroit MI 48201.}

\author{S.~Mak}
\affiliation{Department of Statistical Science, Duke University, Durham NC 27708.}

\author{P.~M.~Jacobs}
\affiliation{Department of Physics, University of California, Berkeley CA 94270.}
\affiliation{Nuclear Science Division, Lawrence Berkeley National Laboratory, Berkeley CA 94270.}

\author{A.~Majumder}
\affiliation{Department of Physics and Astronomy, Wayne State University, Detroit MI 48201.}

\author{A.~Angerami}
\affiliation{Lawrence Livermore National Laboratory, Livermore CA 94550.}

\author{R.~Arora}
\affiliation{Department of Computer Science, Wayne State University, Detroit MI 48202.}

\author{S.~A.~Bass}
\affiliation{Department of Physics, Duke University, Durham, NC 27708, USA}

\author{R.~Datta}
\affiliation{Department of Physics and Astronomy, Wayne State University, Detroit MI 48201.}

\author{L.~Du}
\affiliation{Department of Physics, McGill University, Montr\'{e}al QC H3A\,2T8, Canada.}
\affiliation{Department of Physics, University of California, Berkeley CA 94270.}
\affiliation{Nuclear Science Division, Lawrence Berkeley National Laboratory, Berkeley CA 94270.}

\author{H.~Elfner}
\affiliation{GSI Helmholtzzentrum f\"{u}r Schwerionenforschung, 64291 Darmstadt, Germany.}
\affiliation{Institute for Theoretical Physics, Goethe University, 60438 Frankfurt am Main, Germany.}
\affiliation{Frankfurt Institute for Advanced Studies, 60438 Frankfurt am Main, Germany.}

\author{R.~J.~Fries}
\affiliation{Cyclotron Institute, Texas A\&M University, College Station TX 77843.}
\affiliation{Department of Physics and Astronomy, Texas A\&M University, College Station TX 77843.}

\author{C.~Gale}
\affiliation{Department of Physics, McGill University, Montr\'{e}al QC H3A\,2T8, Canada.}

\author{Y.~He}
\affiliation{Guangdong Provincial Key Laboratory of Nuclear Science, Institute of Quantum Matter, South China Normal University, Guangzhou 510006, China.}
\affiliation{Guangdong-Hong Kong Joint Laboratory of Quantum Matter, Southern Nuclear Science Computing Center, South China Normal University, Guangzhou 510006, China.}

\author{B.~V.~Jacak}
\affiliation{Department of Physics, University of California, Berkeley CA 94270.}
\affiliation{Nuclear Science Division, Lawrence Berkeley National Laboratory, Berkeley CA 94270.}

\author{S.~Jeon}
\affiliation{Department of Physics, McGill University, Montr\'{e}al QC H3A\,2T8, Canada.}

\author{F.~Jonas}
\affiliation{Department of Physics, University of California, Berkeley CA 94270.}
\affiliation{Nuclear Science Division, Lawrence Berkeley National Laboratory, Berkeley CA 94270.}

\author{L.~Kasper}
\affiliation{Department of Physics and Astronomy, Vanderbilt University, Nashville TN 37235.}

\author{M.~Kordell~II}
\affiliation{Cyclotron Institute, Texas A\&M University, College Station TX 77843.}
\affiliation{Department of Physics and Astronomy, Texas A\&M University, College Station TX 77843.}

\author{R.~Kunnawalkam-Elayavalli}
\affiliation{Department of Physics and Astronomy, Vanderbilt University, Nashville TN 37235.}

\author{J.~Latessa}
\affiliation{Department of Computer Science, Wayne State University, Detroit MI 48202.}

\author{Y.-J.~Lee}
\affiliation{Laboratory for Nuclear Science, Massachusetts Institute of Technology, Cambridge MA 02139.}
\affiliation{Department of Physics, Massachusetts Institute of Technology, Cambridge MA 02139.}

\author{R.~Lemmon}
\affiliation{Daresbury Laboratory, Daresbury, Warrington, Cheshire, WA44AD, United Kingdom.}

\author{M.~Luzum}
\affiliation{Instituto  de  F\`{i}sica,  Universidade  de  S\~{a}o  Paulo,  C.P.  66318,  05315-970  S\~{a}o  Paulo,  SP,  Brazil. }

\author{A.~Mankolli}
\affiliation{Department of Physics and Astronomy, Vanderbilt University, Nashville TN 37235.}

\author{C.~Martin}
\affiliation{Department of Physics and Astronomy, University of Tennessee, Knoxville TN 37996.}

\author{H.~Mehryar}
\affiliation{Department of Computer Science, Wayne State University, Detroit MI 48202.}

\author{T.~Mengel}
\affiliation{Department of Physics and Astronomy, University of Tennessee, Knoxville TN 37996.}

\author{C.~Nattrass}
\affiliation{Department of Physics and Astronomy, University of Tennessee, Knoxville TN 37996.}

\author{J.~Norman}
\affiliation{Oliver Lodge Laboratory, University of Liverpool, Liverpool, United Kingdom.}

\author{C.~Parker}
\affiliation{Cyclotron Institute, Texas A\&M University, College Station TX 77843.}
\affiliation{Department of Physics and Astronomy, Texas A\&M University, College Station TX 77843.}

\author{J.-F. Paquet}
\affiliation{Department of Physics and Astronomy, Vanderbilt University, Nashville TN 37235.}

\author{J.~H.~Putschke}
\affiliation{Department of Physics and Astronomy, Wayne State University, Detroit MI 48201.}

\author{H.~Roch}
\affiliation{Department of Physics and Astronomy, Wayne State University, Detroit MI 48201.}

\author{G.~Roland}
\affiliation{Laboratory for Nuclear Science, Massachusetts Institute of Technology, Cambridge MA 02139.}
\affiliation{Department of Physics, Massachusetts Institute of Technology, Cambridge MA 02139.}

\author{B.~Schenke}
\affiliation{Physics Department, Brookhaven National Laboratory, Upton NY 11973.}

\author{L.~Schwiebert}
\affiliation{Department of Computer Science, Wayne State University, Detroit MI 48202.}

\author{A.~Sengupta}
\affiliation{Cyclotron Institute, Texas A\&M University, College Station TX 77843.}
\affiliation{Department of Physics and Astronomy, Texas A\&M University, College Station TX 77843.}

\author{C.~Shen}
\affiliation{Department of Physics and Astronomy, Wayne State University, Detroit MI 48201.}
\affiliation{RIKEN BNL Research Center, Brookhaven National Laboratory, Upton NY 11973.}

\author{M.~Singh}
\affiliation{Department of Physics and Astronomy, Vanderbilt University, Nashville TN 37235.}

\author{C.~Sirimanna}
\affiliation{Department of Physics and Astronomy, Wayne State University, Detroit MI 48201.}
\affiliation{Department of Physics, Duke University, Durham, NC 27708, USA}

\author{D.~Soeder}
\affiliation{Department of Physics, Duke University, Durham NC 27708.}

\author{R.~A.~Soltz}
\affiliation{Department of Physics and Astronomy, Wayne State University, Detroit MI 48201.}
\affiliation{Lawrence Livermore National Laboratory, Livermore CA 94550.}

\author{I.~Soudi}
\affiliation{Department of Physics and Astronomy, Wayne State University, Detroit MI 48201.}
\affiliation{University of Jyväskylä, Department of Physics, P.O. Box 35, FI-40014 University of Jyväskylä, Finland.}
\affiliation{Helsinki Institute of Physics, P.O. Box 64, FI-00014 University of Helsinki, Finland.}

\author{Y.~Tachibana}
\affiliation{Akita International University, Yuwa, Akita-city 010-1292, Japan.}

\author{J.~Velkovska}
\affiliation{Department of Physics and Astronomy, Vanderbilt University, Nashville TN 37235.}

\author{G.~Vujanovic}
\affiliation{Department of Physics, University of Regina, Regina, SK S4S 0A2, Canada.}

\author{X.-N.~Wang}
\affiliation{Key Laboratory of Quark and Lepton Physics (MOE) and Institute of Particle Physics, Central China Normal University, Wuhan 430079, China.}
\affiliation{Department of Physics, University of California, Berkeley CA 94270.}
\affiliation{Nuclear Science Division, Lawrence Berkeley National Laboratory, Berkeley CA 94270.}

\author{X.~Wu}
\affiliation{Department of Physics, McGill University, Montr\'{e}al QC H3A\,2T8, Canada.}
\affiliation{Department of Physics and Astronomy, Wayne State University, Detroit MI 48201.}

\author{W.~Zhao}
\affiliation{Department of Physics and Astronomy, Wayne State University, Detroit MI 48201.}
\affiliation{Department of Physics, University of California, Berkeley CA 94270.}
\affiliation{Nuclear Science Division, Lawrence Berkeley National Laboratory, Berkeley CA 94270.}

\collaboration{The JETSCAPE Collaboration}

\date{\today}

\begin{abstract}

The \JETSCAPE\ Collaboration reports a new determination of the jet transport parameter \qhat\ in the Quark-Gluon Plasma (QGP) using Bayesian Inference, incorporating all available inclusive hadron and jet yield suppression data measured in heavy-ion collisions at RHIC and the LHC. This multi-observable analysis extends the previously published \JETSCAPE\ Bayesian Inference determination of \qhat, which was based solely on a selection of inclusive hadron suppression data. \JETSCAPE\ is a modular framework incorporating detailed dynamical models of QGP formation and evolution, and jet propagation and interaction in the QGP. Virtuality-dependent partonic energy loss in the QGP is modeled as a thermalized weakly-coupled plasma, with parameters determined from Bayesian calibration using soft-sector observables. This Bayesian calibration of \qhat\ utilizes Active Learning, a machine--learning approach, for efficient exploitation of computing resources. The experimental data included in this analysis span a broad range in collision energy and centrality, and in transverse momentum. 
In order to explore the systematic dependence of the extracted parameter posterior distributions, several different calibrations are reported, based on combined jet and hadron data; on jet or hadron data separately; and on restricted kinematic or centrality ranges of the jet and hadron data. Tension is observed in comparison of these variations, providing new insights into the physics of jet transport in the QGP and its theoretical formulation.

\end{abstract}

\maketitle

\setcounter{page}{2}

\input{Introduction}

\input{PhysicsModel} 
\input{BayesianInference}

\input{Results}

\input{Summary}

\section*{Acknowledgements}
This work was supported in part by the National Science Foundation (NSF) within the framework of the JETSCAPE collaboration, under grant number OAC-2004571 (CSSI:X-SCAPE). It was also supported under  PHY-1516590 and PHY-1812431 (R.J.F., M.Ko., C.P. and A.S.), by PHY-2012922 (C.Sh.); it was supported in part by the US Department of Energy, Office of Science, Office of Nuclear Physics under grant numbers \rm{DE-AC02-05CH11231} (B.J., P.M.J., X.-N.W., and W.Z.), \rm{DE-AC52-07NA27344} (A.A., R.A.S.), \rm{DE-SC0013460} (A.K., A.M., C.Sh., I.S., C.Si, R.D. and X.W.), \rm{DE-SC0021969} (C.Sh. and W.Z.), \rm{DE-SC0012704} (B.S.), \rm{DE-FG02-92ER40713} (J.H.P.), \rm{DE-FG02-05ER41367} (D.S. and S.A.B.), \rm{DE-SC0024660} (R.K.E), \rm{DE-SC0024347} (J.-F.P. and M.S.). The work was also supported in part by the National Science Foundation of China (NSFC) under grant numbers 11935007, 11861131009 and 11890714 (Y.H. and X.-N.W.), by the Natural Sciences and Engineering Research Council of Canada (C.G., S.J., and G.V.),  by the University of Regina President's Tri-Agency Grant Support Program (G.V.), by the Canada Research Chair program (G.V. and A.K.) reference number CRC-2022-00146, by the Office of the Vice President for Research (OVPR) at Wayne State University (Y.T.), and by the S\~{a}o Paulo Research Foundation (FAPESP) under projects 2016/24029-6, 2017/05685-2 and 2018/24720-6 (M.L.). 
I.~S. was funded as a part of the European Research Council project ERC-2018-ADG-835105 YoctoLHC , and as a part of the Center of Excellence in Quark Matter of the Academy of Finland (project 346325).
C.Sh., J.-F.P. and R.K.E. acknowledge a DOE Office of Science Early Career Award. 

A portion of the computations related to the bulk were carried out on the National Energy Research Scientific Computing Center (NERSC), a U.S. Department of Energy Office of Science User Facility operated under Contract No. DE-AC02-05CH11231. The bulk medium simulations were done using resources provided by the Open Science Grid (OSG) \cite{Pordes:2007zzb, Sfiligoi:2009cct}, which is supported by the National Science Foundation award \#2030508. Data storage was provided in part by the OSIRIS project supported by the National Science Foundation under grant number OAC-1541335.
Jet quenching calculations were carried out at 
the Expanse system at the San Diego Supercomputer Center (SDSC),
the Bridges-2 system, which is supported by National Science Foundation award number ACI-1928147, at the Pittsburgh Supercomputing Center (PSC),
and the Stampede2 system at Texas Advanced Computing Center (TACC), The University of Texas at Austin,
through allocation PHY200093 from the Extreme Science and Engineering Discovery Environment (XSEDE), which was supported by NSF grant number \#1548562.

\input{Appendix}

\clearpage{}%

\bibliographystyle{utphys}   
\bibliography{bibliography}

\end{document}

%% file: commands.tex
\usepackage{xcolor}
\definecolor{myOrange}{rgb}{1,0.5,0.}
\definecolor{myGreen}{rgb}{0.0,0.6,0.1}

\newcommand{\pp}           {pp\xspace}

\newcommand{\PbPb}         {\mbox{Pb--Pb}\xspace}

\newcommand{\AuAu}         {\mbox{Au--Au}\xspace}

\newcommand{\snn}          {\ensuremath{\sqrt{s_{\mathrm{NN}}}}\xspace}

\newcommand{\RAA}          {\ensuremath{R_{\mathrm{AA}}}\xspace}

\newcommand{\nineH}        {$\sqrt{s}~=~0.9$~Te\kern-.1emV\xspace}
\newcommand{\seven}        {$\sqrt{s}~=~7$~Te\kern-.1emV\xspace}
\newcommand{\twoH}         {$\sqrt{s}~=~0.2$~Te\kern-.1emV\xspace}
\newcommand{\twosevensix}  {$\sqrt{s}~=~2.76$~Te\kern-.1emV\xspace}
\newcommand{\five}         {$\sqrt{s}~=~5.02$~Te\kern-.1emV\xspace}
\newcommand{\twosevensixnn}{$\sqrt{s_{\mathrm{NN}}}~=~2.76$~Te\kern-.1emV\xspace}
\newcommand{\fivenn}       {$\sqrt{s_{\mathrm{NN}}}~=~5.02$~Te\kern-.1emV\xspace}

\newcommand{\GeVc}         {Ge\kern-.1emV/$c$\xspace}
\newcommand{\MeVc}         {Me\kern-.1emV/$c$\xspace}
\newcommand{\TeV}          {Te\kern-.1emV\xspace}
\newcommand{\GeV}          {Ge\kern-.1emV\xspace}
\newcommand{\MeV}          {Me\kern-.1emV\xspace}
\newcommand{\GeVmass}      {Ge\kern-.2emV/$c^2$\xspace}
\newcommand{\MeVmass}      {Me\kern-.2emV/$c^2$\xspace}

\newcommand{\lQCD}{\ensuremath{\Lambda_{\mathrm{QCD}}}}

\newcommand{\TC}{\ensuremath{T_{\mathrm{C}}}}
\newcommand{\qsqr}{\ensuremath{Q^2}}

\newcommand{\qhat}{\ensuremath{\hat{q}}\xspace}
\newcommand{\qhatTcubed}{\ensuremath{\hat{q}/{T}^3}\xspace}

\newcommand{\alphas}{\ensuremath{\alpha_{\text{s}}}\xspace}
\newcommand{\alphasrun}{\ensuremath{\alpha_{\text{s}}^\mathrm{run}}\xspace}
\newcommand{\alphasfix}{\ensuremath{\alpha_{\text{s}}^\mathrm{fix}}\xspace}

\newcommand{\qswitch}{\ensuremath{Q_{\text{0}}}\xspace}
\newcommand{\tstart}{\ensuremath{\tau_{\text{0}}}\xspace}

\newcommand{\aaa}{\ensuremath{\mathrm{A+A}}}
\newcommand{\sqrtsNN}{\ensuremath{\sqrt{s_\mathrm{NN}}}}
\newcommand{\sqrts}{\ensuremath{\sqrt{s}}}

\newcommand{\pT}{\ensuremath{p_\mathrm{T}}}
\newcommand{\pTjet}{\ensuremath{p_\mathrm{T, jet}}}

\newcommand{\gev}{\ensuremath{\mathrm{GeV/}c}}
\newcommand{\tev}{\ensuremath{\mathrm{TeV}}}

\newcommand{\rr}{\ensuremath{R}}

\newcommand{\Eref}{\ensuremath{E_\mathrm{ref}}}
\newcommand{\Tref}{\ensuremath{T_\mathrm{ref}}}

\newcommand{\JETSCAPE}{\textsc{Jetscape}}
\newcommand{\AMY}{\textsc{Amy}}

\newcommand{\URQMD}{\textsc{UrQMD}}
\newcommand{\PYTHIA}{\textsc{Pythia}}

\newcommand{\MATTER}{\textsc{Matter}}
\newcommand{\LBT}{\textsc{Lbt}}
\newcommand{\LBTtwo}{\textsc{Lbt2}}
\newcommand{\TRENTO}{\textsc{Trento}}

\newcommand{\JET}{\textsc{Jet}}

%% file: Introduction.tex
\section{Introduction}
\label{sect:Intro}

Strongly-interacting matter at high energy density forms a deconfined Quark-Gluon Plasma~\cite{Busza:2018rrf,Harris:2023tti}. Numerical calculations using Quantum Chromodynamics (QCD) on a lattice predict that equilibrated matter with zero baryo-chemical potential has a cross-over transition to a QGP at temperature $\TC\approx{155}$ MeV~\cite{Borsanyi:2013bia,Bhattacharya:2014ara,HotQCD:2014kol,Ding:2015ona,Bazavov:2017dus}. However, the deconfined QGP has fewer degrees of freedom than the Stefan-Boltzmann limit of a non-interacting  gas of partons (quarks and gluons) up to much higher temperature than \TC, indicating that its constituents have significant interactions.

The QGP filled the early universe a few micro-seconds after the Big Bang, and it is recreated today in energetic collisions of heavy atomic nuclei at the Large Hadron Collider (LHC) at CERN and the Relativistic Heavy-Ion Collider (RHIC) at Brookhaven National Laboratory~\cite{Busza:2018rrf,Harris:2023tti}. Experimental measurements at these facilities, and their comparison to theoretical calculations, show that the QGP flows as a fluid with very small specific shear viscosity~\cite{Heinz:2013th,JETSCAPE:2020mzn,JETSCAPE:2020shq,Nijs:2020roc} and is opaque to the passage of energetic color charges~\cite{Majumder:2010qh,Cunqueiro:2021wls,Apolinario:2022vzg}.

Jets in hadronic collisions arise from the hard (high momentum--transfer \qsqr) interaction of partons from the projectiles. The scattered partons are initially highly virtual, coming on-shell by radiation of a gluon shower which manifests in a collimated spray of hadrons that is  observable experimentally. Jet production and jet structure have been measured extensively at colliders, with high-order perturbative QCD (pQCD) calculations in excellent agreement with the data~\cite{Abelev:2006uq,Adamczyk:2016okk,Abelev:2013fn,Aad:2014vwa,Khachatryan:2016mlc,CMS:2016jip,ATLAS:2017ble,Acharya:2019jyg,ATLAS:2011myc,ALICE:2018ype,ATLAS:2019rqw,ALICE:2020pga}. 

In nuclear collisions, hard-scattered partons are generated prior to formation of the QGP and interact with it; such interactions modify observed jet production rates and jet structure relative to those in vacuum (``jet quenching'') ~\cite{Majumder:2010qh,Cunqueiro:2021wls,Apolinario:2022vzg}. Of especial note is energy loss due to jet quenching, in which jet--QGP interactions induce energy transfer away from the hardest jet shower branch for inclusive hadron production, or outside of the jet cone for reconstructed jets. Such energy loss is often measured as \RAA, which is the ratio of inclusive hadron or jet production yields at the same transverse momentum (\pT) in \aaa\ and \pp\ collisions, scaled to account for nuclear geometric effects~\cite{Cunqueiro:2021wls}. A value $\RAA\sim1$ indicates negligible jet quenching effects, whereas $\RAA\ll1$ (i.e. inclusive yield suppression) indicates significant energy loss due to jet quenching.

Jet quenching effects are likewise calculable theoretically, and comparisons of jet quenching calculations and measurements provide unique probes of the structure and dynamics of the QGP~\cite{Majumder:2010qh,Armesto:2011ht,Cunqueiro:2021wls,Apolinario:2022vzg}. Jet quenching calculations incorporate elastic and radiative interactions of the jet-initiating parton and its gluon shower with the QGP. Various theoretical frameworks have been developed for calculating radiative interactions, using different approximation schemes: \AMY\, based on a Hard Thermal Loop approach~\cite{{Arnold:2001ms,Arnold:2002ja}}; BDMPS, which uses a soft radiation approximation
~\cite{Baier:1996sk,Baier:1998kq,Zakharov:1996fv,Zakharov:1997uu,Wiedemann:2000za,Salgado:2003gb,Armesto:2003jh}; GLV, which utilizes an opacity expansion~\cite{Gyulassy:1999zd,Gyulassy:2000er,Gyulassy:2001nm}; and Higher-Twist~\cite{Wang:2001ifa,Majumder:2009ge}. Comparison of different theoretical formulations of \qhat is presented in Ref.~\cite{Armesto:2011ht}. Jet quenching models may also include the QGP response to jet energy dissipated in the interaction~\cite{Cao:2020wlm,KunnawalkamElayavalli:2017hxo}.

Comparison of jet quenching calculations with experimental data has been used to constrain the QGP jet transport coefficient \qhat~\cite{Majumder:2010qh,Cunqueiro:2021wls,Apolinario:2022vzg}, which characterizes the momentum transfer between an energetic partonic probe and the QGP. These constraints are commonly expressed in terms of the distribution of \qhatTcubed, where $T$ is the QGP temperature, in order to factor out the expected leading $T$ dependence of \qhat\ (the density of scattering centers in a thermal medium varies approximately as ${T}^3$). Current constraints on \qhatTcubed\ incorporate only a limited subset of available jet quenching measurements: hadron \RAA~\cite{Andres:2016iys,Feal:2019xfl,Burke:2013yra,JETSCAPE:2021ehl,Xie:2020zdb}; hadron \RAA, and di--hadron and $\gamma$--hadron correlations~\cite{Xie:2022ght}; or selected hadron and jet \RAA~\cite{Ke:2020clc}; these different analyses generate inconsistent constraints, however, due to differing modelling assumptions and approximations, and different data selection~\cite{Apolinario:2022vzg}. A key question is to determine whether \qhatTcubed\ is a universal property of the QGP, whose extracted distribution is independent of how it is probed.

This paper presents a new determination of \qhatTcubed\ using the multi-stage \JETSCAPE\ framework~\cite{Putschke:2019yrg}, which incorporates a detailed 2+1 D hydrodynamic model with parameters determined by Bayesian calibration of soft observables~\cite{Bernhard:2019bmu}, and with virtuality-dependent jet quenching calculated using the \MATTER~\cite{Majumder:2013re,Cao:2017qpx} and \LBT~\cite{Cao:2017hhk,Chen:2017zte,Luo:2018pto,JETSCAPE:2017eso} models. Constraints on \qhatTcubed\ are determined by Bayesian Inference, incorporating all inclusive hadron and inclusive jet \RAA\ measurements for central and semi--central \aaa\ collisions at the LHC and RHIC published prior to February 2022. This Bayesian calibration is computationally expensive, however, requiring simulations spanning a large parameter space, and a machine-learning based approach, called Active Learning~\cite{cohn1996active,mak2017information,chen2022adaptive,song2023ace}, is utilized for efficient exploitation of computing resources.

Consistency of \qhatTcubed\ posterior distributions extracted solely from hadron or jet \RAA, from different kinematic ranges, and from different centrality intervals, is explored for the first time. This study demonstrates the discriminating power of such a multi-observable approach, and points towards yet broader future studies.

The paper is structured as follows: Sect. \ref{sec:physicsModel} presents the \JETSCAPE\ physics;
Sect. \ref{sec:simulationsAndInferece} presents details of the simulations, the experimental measurements used in the analysis, and the subsequent Bayesian Inference; Sect. \ref{sec:results} presents the analysis results, including differential comparisons; and Sect.~\ref{sect:Summary} presents the summary and conclusions.

%% file: PhysicsModel.tex
\section{Physics model}
\label{sec:physicsModel}

\subsection{Factorization}
\label{sect:factorization}

Analytic QCD calculations of high \qsqr\ processes are based on factorization, i.e. division into sub-processes at different momentum scales, each of which is characterized by a probability distribution that does not depend on other sub-processes and whose interface is characterized by a few parameters. Factorization is applicable for inclusive jet production in \pp\ collisions at high energy, whose factorized cross section is written~\cite{Collins:1983ju,Collins:1985ue,Collins:1988ig,Collins:1989gx}:
\begin{eqnarray}
    \frac{d^3 \sigma}{dy d^2p_T} &=& \int d x_a d x_b G(x_a,\mu^2) G(x_b,\mu^2) \frac{d \hat{\sigma}}{d\hat{t}} \frac{J(z,\mu^2)}{\pi z}, \label{eq:factorized}
\end{eqnarray}

\noindent
where $G(x, \mu^2) $ is the parton distribution function (PDF) for a parton carrying a fraction $x$ of the forward light-cone momentum of the proton; $\hat{\sigma}$ is the partonic cross section as a function of Mandelstam variable $\hat{t}$; $J(z,\mu^2)$ is the jet function which specifies the multiplicity of jets carrying a momentum fraction $z$ of the forward momentum; and $\mu \gg \Lambda_\mathrm{QCD}$ is the scale at which $G$ and $J$ are evaluated. The inclusive hadron production cross section is obtained by replacing $J(z,\mu^2)$ by the Fragmentation Function $D(z,\mu^2)$, with the hadron carrying the momentum fraction $z$. Evaluation of these cross sections can also be carried out using Monte Carlo event generators which produce multi-particle states that model the events recorded by collider experiments.

Model calculations in this analysis are carried out using the \JETSCAPE\ framework~\cite{Putschke:2019yrg}, in which distinct simulation modules calculate each independent element of the factorized process and are combined sequentially for the simulation of complete \pp\ or \aaa\ collisions. The initial state radiation and hard scattering are simulated using \PYTHIA\ (8.235 default tune)~\cite{Sjostrand:2014zea} with Final State Radiation (FSR) turned off. Final state radiation in \pp\ collisions is calculated using the \MATTER\ generator~\cite{Majumder:2013re, Cao:2017qpx}. A separate instance of the \PYTHIA\ generator is used for hadronization of hard processes. See Ref.~\cite{JETSCAPE:2019udz} for calculational details, comparison with \pp\ data, and the full parameters of the JETSCAPE PP19 tune parameters used in this analysis.

\subsection{Simulation of \aaa\ collisions}
\label{sect:AAsimulations}

The calculation of jet interactions in the QGP generated in \aaa\ collisions requires simulation of the creation and evolution of bulk matter which is calculated first, followed by simulation of hard parton showers which propagate in the evolving bulk medium. For the bulk simulation, the initial nucleon and energy distributions are calculated using the \TRENTO\ model~\cite{Moreland:2014oya}. The system initially evolves by free streaming \cite{Liu:2015nwa} for a period $\tau_R$,  followed by a viscous fluid dynamic stage corresponding to the expansion and cooling of the hot QGP that is simulated by Israel-Stewart transient hydrodynamics, as implemented in the VISHNU code~\cite{Song:2007ux,Shen:2014vra}. When an element of the QGP cools below a switching  temperature $T_{SW}$, it is hadronized using the  the Cooper-Frye approach~\cite{Cooper:1974mv}. Subsequent multiple scattering of hadrons is simulated using the \URQMD\ model~\cite{Bass:1998ca,Bleicher:1999xi}. The bulk matter was simulated with parameters corresponding to the maximum a posteriori determined in Ref.~\cite{Bernhard:2019bmu}, which produces similar bulk properties to those obtained in a previous \JETSCAPE\ Bayesian analysis~\cite{JETSCAPE:2020mzn,JETSCAPE:2020shq}. The existing profiles are utilized for expediency. These parameters include those of the \TRENTO\ model of initial conditions; the time of transition from free streaming to hydrodynamics; parameters specifying the temperature dependence of the shear and bulk viscosities; and parameters of bulk hadronization. See Ref.~\cite{JETSCAPE:2022jer} for details.

For \aaa\ collisions, the geometric distribution of hard-scattering processes within the QGP is determined by sampling the distribution of nucleon-nucleon collisions generated by \TRENTO. Initial--state radiation prior to the hard scattering and the hard scattering itself are modeled using the \PYTHIA\ generator, as described above for \pp\ collisions. Final--state radiation, both in--vacuum and in--medium, is simulated by other Monte Carlo models, as described below.
The off-shellness or virtuality of a parton generated in a hard interaction is typically of the order of (though smaller than) its energy. In vacuum, a hard parton decays in a cascade of progressively lower--energy and lower--virtuality partons. This process continues until the virtuality of the partons reaches a scale at which interactions are non-perturbative. While the partonic cascade at large scales $(\mu \gg \Lambda_\mathrm{QCD})$ can be described using perturbation theory, the soft stage of the shower must be treated non-pertubatively, typically by a hadronization model. In this paper, hard sector hadronization will be carried out as in Ref.~\cite{JETSCAPE:2022jer}, and none of the parameters of hadronization will be included in the Bayesian analysis. 

\subsection{Specification of \qhat}
\label{sect:qhat}

The presence of a hot, dense medium modifies this cascade-like decay of a hard virtual parton, corresponding to jet quenching. In this case, partons in the developing cascade scatter from constituents in the medium, leading to the emission of more partons, with consequent re-distribution of the energy of the shower to wider angles than in vacuum. This process is characterized by the transport coefficient \qhat, which is the mean square transverse momentum exchanged between a parton and the medium per unit length traversed,

\begin{align}
    \qhat &= \frac{\langle k_\perp^2 \rangle_L }{L} \xrightarrow[{\rm Scatterings}]{}\frac{1}{L} \left\langle \left( \sum_{j}^{ N_\mathrm{scat} }  {\vec{k}_{{\perp},j}} \right)^{2} \right\rangle \nonumber \\
    &\xrightarrow[\lambda_\mathrm{corr}<\lambda_\mathrm{mfp}]{} \frac{1}{L} \sum_{j}^{ N_\mathrm{scat} } \langle {k_\perp}_j^2 \rangle . 
    \label{eq:qhat-def}
\end{align}

In this equation, the second expression applies if the transverse momentum exchange can be decomposed into separate scatterings with constituents in the medium, where $N_\mathrm{scat}$ is the number of scatterings in a length $L$. The notation $\langle \cdots \rangle$ indicates averaging over an ensemble of events used in a jet quenching measurement or a Monte Carlo simulation, or the sum over theoretical configurations in a semi-analytical or lattice calculation. Different choices of $L$ yield different averages over the medium. For a static medium the choice of $L$ is largely irrelevant, since \qhat is the same everywhere. For a dynamically evolving medium, a large value of $L$ will average over a range of temperatures, while a very short value of $L$ will approach a local quantity but requires many events (configurations) to produce sufficient statistics. The third expression applies if the in-medium correlation length $\lambda_\mathrm{corr}$ is less than the scattering mean free path $\lambda_\mathrm{mfp}$, such that successive scatterings do not interfere significantly. This is typically assumed in jet quenching calculations, reducing the contribution to \qhat from $N_\mathrm{scat}$ multiple scatterings to the simple sum of $N_\mathrm{scat}$ single scatterings.

\begin{figure}
    \centering
    \includegraphics[width=0.45\textwidth]{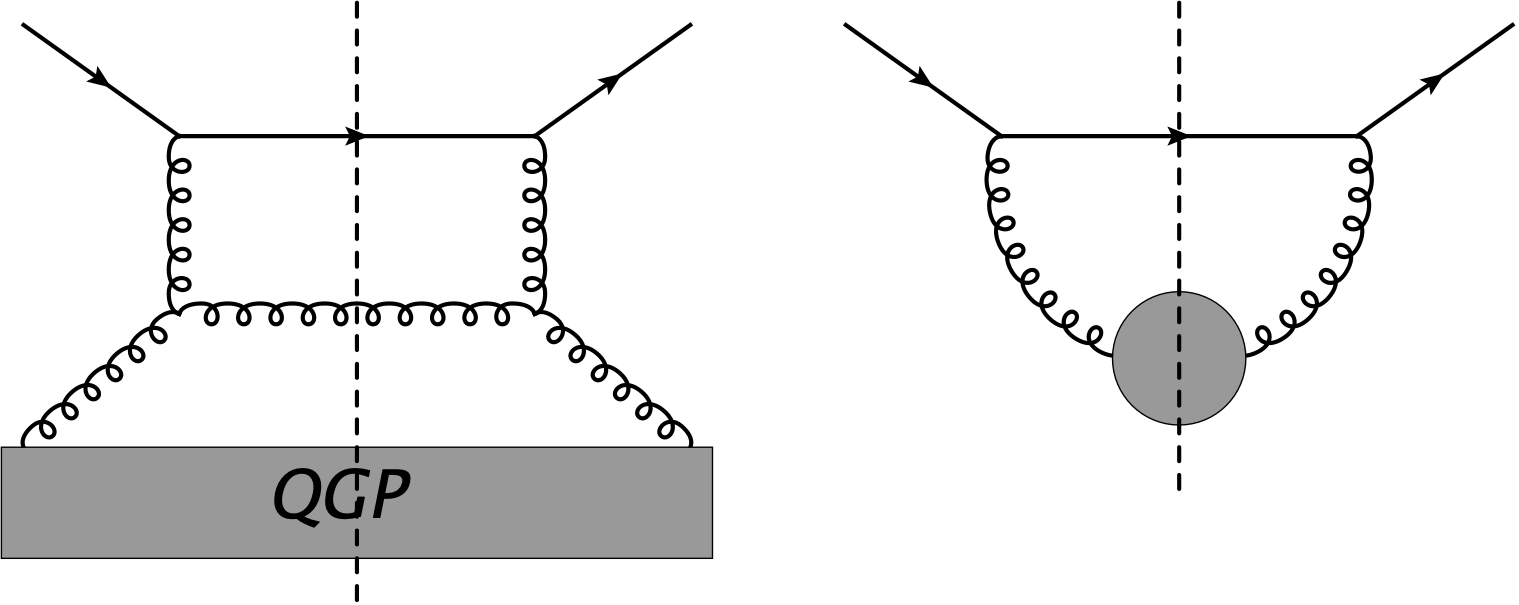}
    \caption{Left: Single scattering of a hard parton with a gluon in the Quark-Gluon Plasma. In this diagram the hard parton is a quark and the medium parton is a gluon, but calculations also include hard gluons and quarks in the QGP. Right: The same process described in HTL effective theory, which is used to calculate \qhat. }
    \label{fig:scattering-diagram}
\end{figure}

In a field--theoretic calculation of a parton scattering off a gluon field in a locally thermalized medium with temperature $T$, the single-scattering limit of \qhat corresponds to the Fourier transform of the correlation of gluon field strength tensors, 
\begin{align}
\hat{q} &=   \frac{16 \pi \alpha_s \sqrt{2} C_R }{ ( N_{c}^2 -1 ) }  \int \frac{ dy^{-} d^{2} y_{\perp} } {(2\pi)^3}  d^{2} k_{\perp}  e^{ -i  \frac{\vec{k}^{2}_{\perp}}{2q^{-}} y^{-} +i\vec{k}_{\perp}.\vec{y}_{\perp} }  \nonumber \\
& \times \sum _{n} \langle n | \frac{e^{-E_{n}/T}}{Z}  \mathrm{Tr}[F^{+ j}(0) F^{+}_{j}(y^{-},y_{\perp})]  | n \rangle, \label{eq:qhat-general}
\end{align}
\noindent
where $C_R$ is the representation--specific Casimir factor (for a quark, $C_{R}=C_F=(N_C^2-1)/(2N_C)$), and \alphas is the strong coupling constant at the scattering vertex of the hard quark and gluon field. States of the ensemble are represented by $|n\rangle$, and $F^{ \mu\nu} = t^{ a} F^{ a\mu\nu}$ is the bare gauge field-strength tensor (with $j=1,2$ denoting transverse directions). The use of gauge links on the two field strength tensors, as in Ref.~\cite{Benzke:2012sz}, will render the above expression gauge invariant. 
If the states $|n \rangle$ are replaced with a plasma of quarks and gluons and weak coupling is assumed to be applicable, at least for the purposes of jet modification, then the expectation in Eq.~\eqref{eq:qhat-general} can be calculated analytically .

Figure~\ref{fig:scattering-diagram} shows the single scattering process in the QGP experienced by a hard parton, which modifies its evolving shower. The left diagram shows a hard quark scattering from a gluon in the QGP. The right diagram shows this process as implemented in Hard Thermal Loop (HTL) effective theory~\cite{Braaten:1989mz,Frenkel:1989br}, which assumes that the QGP is a thermalized, weakly--coupled QCD plasma. The gray circle is the HTL self-energy. In this approach, \qhat\ is expressed to Leading Order (LO) in the coupling as
\begin{eqnarray}
     \frac{{\qhat}^{HTL}}{T^3}  = C_{a}\frac{50.48}{\pi} \alphasrun \alphasfix  \log \left[\frac{2ET}{6\pi T^2 \alphasfix} \right]. 
    \label{eq:HTL-q-hat}
\end{eqnarray}
\noindent
Next-to-leading-order (NLO) corrections to Eq.~\ref{eq:HTL-q-hat} have been calculated~\cite{Caron-Huot:2008zna}, with the perturbative series found to have poor convergence properties. This issue is discussed further in Sect.~\ref{sec:results}.

The scaling of \qhat\ by $1/T^3$ in Eq.~\ref{eq:HTL-q-hat} factors out its leading $T$ dependence, as discussed above, and \qhatTcubed\ is the expression extracted in the Bayesian Inference analysis. The functional dependence of \qhatTcubed\ on $E/T$ is specified, while the specific contributions of $T$ to the numerator and denominator of the logarithmic argument are also shown separately to clarify their origin. The temperature $T$ varies spatially and temporally within the evolving QGP, and in simulations its local value is determined from the pre-calibrated simulation of the bulk medium. 

Eq.~\ref{eq:HTL-q-hat} incorporates two different values of the strong coupling \alphas, denoted as \alphasrun\ and \alphasfix. This is a consequence of letting all couplings run from the thermal and Debye scales $T, m_D \simeq gT \lesssim 1$~GeV, up to the hard scale $\bar{\mu} \approx \sqrt{2ET}$, where $E$ is the energy of the jet parton. The value of \alphasfix\ is taken as the coupling at the soft thermal Debye scale, and is referred to below simply  as \alphas. The coupling at the hard scale ($\bar{\mu} \simeq \sqrt{2 ET}$), \alphasrun\ is defined as 

\begin{equation}
    \alphasrun (\mu^2) = \frac{12\pi}{[ 11N_c - 2N_f ]} \frac{1}{\log \left( \frac{\mu^2}{\Lambda^2} \right) },
\end{equation}
\noindent
where $\Lambda$ is chosen such that $\alphasrun (\mu^2 \leq 1\mathrm{GeV}^2) = \alphasfix$. 

Multiple scattering of a hard parton in a medium raises its virtuality. For parton lifetime $\tau$, the total increase in virtuality is estimated to be~\cite{Baier:1996sk}:
\begin{eqnarray}
    \mu_{\rm med}^2 \simeq \hat{q} \tau. 
\end{eqnarray}
For parton virtuality much larger than $\mu_{\rm med}$, emissions are mostly vacuum--like, with minor modification from rare scatterings in the medium (denoted the ``high--virtuality'' stage). As the virtuality of cascading partons approaches $\mu_{\rm med}$, they enter the stage of multiple scatterings per emission with a significant increase in the magnitude of energy loss (denoted the ``low--virtuality'' stage). In this analysis, simulations at high virtuality stage are carried out using the \MATTER\ generator~\cite{Majumder:2013re,Cao:2017qpx}, while simulations at low virtuality stage are carried out using the \LBT\ generator~\cite{He:2015pra}.

Equation~\eqref{eq:HTL-q-hat} specifies \qhat\ for both the \MATTER\ model (high virtuality) and \LBT\ model (low virtuality). The diagrams in Fig.~\ref{fig:scattering-diagram} are used to calculate both \qhat\ and the distribution of recoil partons. The outgoing gluon (cut gluon line in the left diagram) is a recoil parton which is tracked by the simulation and is included in the calculation of fragmentation into jet hadrons. 

In the high--virtuality stage, the small transverse size of the highly virtual radiating antenna causes typical scatterings to not resolve the antenna, which is referred to as the ``coherence effect" in jet quenching~\cite{Mehtar-Tani:2011hma,Casalderrey-Solana:2011ule}. This effectively reduces the interaction between the medium and the hard partons. However, in the implementation of this effect there is usually no interaction in the high virtuality phase, in which case the evolving shower undergoes a sudden shift from no interaction to full interaction as it transitions from the high to low--virtuality stage~\cite{Caucal:2018dla}.

In contrast, this analysis utilizes an implementation of coherence in which interaction with the medium increases with decreasing virtuality of the hard parton~\cite{Kumar:2019uvu,Cao:2021rpv,JETSCAPE:2022jer}. This goes beyond the typical reduction of the medium-induced emission kernel at high virtuality. The medium--induced portion of the gluon emission kernel is suppressed by the square of the hard scale. This can be seen in the gluon emission rate from at most one rescattering, expressed for gluons carrying a lightcone momentum $q^+ y$, and transverse momentum $\mu\sqrt{y(1-y)}$ from a quark with lightcone momentum $q^+$ (using the higher-twist formalism) as,  
\begin{eqnarray}
    \frac{dN_g}{dyd\mu^2} &=& \frac{\alphas(\mu^2)}{2 \pi \mu^2 } P(y) \left[1 + \int d\xi^+ \frac{\qhat(\mu^2)}{\mu^2 y (1-y) } \right. \nonumber \\ 
    &\times& \left.  \left\{ 2 - 2 \cos\left( \frac{\mu^2}{2q^+} \xi^+ \right)\right\} \right].  
\end{eqnarray}
The length integral $\xi^+$ is carried out from the origin of a parton to its formation length $\tau = 2q^+/\mu^2$, at which point it will decay by radiating a gluon. The term with value unity in the square brackets represents the vacuum like contribution. The factor $P(y)$ is the unregulated vacuum splitting function. 

At large virtuality $\mu^2$ the second term in square brackets is negligible, with its contribution growing as $\mu^2$ approaches $\hat{q} \tau$. At $\mu^2 \simeq \hat{q} \tau$ the second term is as large as the first term, at which point the parton is transitioned to the lower virtuality stage of the calculation that only includes medium--induced contributions to parton splitting. In the simulations presented below this transition occurs at the scale \qswitch, which is a parameter of the model: partons with virtuality $\mu > \qswitch$ are treated as high virtuality, while partons with $\mu \leq Q_0$ are treated as low virtuality. 

In addition to its $1/\mu^2$ dependence, $\hat{q}$ is also reduced by coherence effects at higher virtuality. 
In Ref.~\cite{Kumar:2019uvu}, the weakening of the interaction between the jet and the medium with increasing virtuality is derived using the effective parton distribution function (PDF) of an incoming parton from the QGP (Fig.~\ref{fig:scattering-diagram}). The weakening of the interaction with the medium is parametrized in this study as 
\begin{align}
\qhat(\mu^2) &= f(\mu^2) {\qhat}^{HTL} \nonumber \\
f(\mu^2 ) &= N \frac{e^{c_3 \left( 1 - \frac{\mu^2}{2ME} \right)} - 1 }{ 1 + c_1 \log \left( \frac{\mu^2}{\lQCD^2}\right) + c_2 \log^2 \left( \frac{\mu^2}{\lQCD^2}\right)  },
\label{Eq:fmu}    
\end{align}
where $N = 1/f(Q_0^2)$; in the low virtuality phase this reduces to $\hat{q}^{\text{HTL}}$. The value of $M$ is taken to be the proton mass. The parameter $c_3$ governs the value of \qhat\ at large $\mu^2$, and influences its evolution as a function of $\mu^2$. The same factor $f$ is also multiplied with the scattering cross section between hard and recoiling medium parton (Fig.~\ref{fig:scattering-diagram}), so that the recoil distribution is consistent with the effective value of $\hat{q}$.

\subsection{Comparison to other \qhat\ formulations}
\label{sect:qhatold}

Eq.~\eqref{eq:qhat-general} shows that, in the single-scattering limit, \qhat is not entirely an intrinsic property of the medium. Rather, it depends on the medium temperature through the thermal partition weight. It also depends on the parton energy $q^-$, and the hard--parton flavor (quark or gluon) via the Casimir factor. Operator products, such as the product of field strength tensors, must be renormalized, thereby becoming scheme and scale--dependent. Thus, in this framework, \qhat is a scheme--dependent quantity which depends on all relevant scales in the calculation, including the medium temperature, and the energy and virtuality of the hard parton. Its calculation and extraction from comparison with data will also depend on the dynamical modeling of the thermal medium that is employed.

A comparison of different theoretical formalisms to describe jet quenching was given in Ref.~\cite{Armesto:2011ht}. Elucidation of the differences of these formulations in practice requires comparison of their quantitative constraints on \qhatTcubed, whose current status is presented in Ref.~\cite{Apolinario:2022vzg}. This section recalls for reference the formalism used in the previous \JETSCAPE\ Bayesian calibration of \qhatTcubed~\cite{JETSCAPE:2021ehl}, to which we compare the the current analysis in Sect.~\ref{sect:Compareqhatold}.

The jet quenching calculation in Ref.~\cite{JETSCAPE:2021ehl} likewise applies a multi-stage approach as a function of parton virtuality, utilizing \MATTER\ for the high--virtuality phase and \LBT\ for the low--virtuality phase. However, its most significant difference from the current analysis is its treatment of the virtuality dependence of \qhatTcubed: rather than a continuous decrease with increasing vituality $\mu^2$ (Eq.~\ref{Eq:fmu}), \qhatTcubed actually increases with increasing virtuality in the high virtuality \MATTER\ stage.

We focus here specifically on the  ``\MATTER\ + \LBTtwo'' parametrization~\cite{JETSCAPE:2021ehl}, whose results are compared to those of the current analysis in Sect.~\ref{sec:results}: 

\begin{widetext}
\begin{equation}
\frac{\qhat\left(\mu,E,T\right) |_{\qswitch,A,C,D}}{T^3}=42C_R\frac{\zeta(3)}{\pi}\left(\frac{4\pi}{9}\right)^2\left\{\frac{A\left[\log\left(\frac{\mu}{\Lambda}\right)-\log\left(\frac{\qswitch}{\Lambda}\right)\right]}{\left[\log\left(\frac{\mu}{\Lambda}\right)\right]^2}\theta (\mu-\qswitch)+\frac{C\left[\log\left(\frac{E}{T}\right)-\log(D)\right]}{\left[\log\left(\frac{ET}{\Lambda^2}\right)\right]^2}\right\}.
\label{eq:MatterLBTtwo}
\end{equation}
\end{widetext}

This formulation has four parameters: \qswitch, $A, C,$ and $D$. The first term in Eq.~\ref{eq:MatterLBTtwo} is dependent only on the parton virtuality $\mu$, but not the medium temperature $T$; it is sensitive to the high--virtuality phase and is largely driven by the \MATTER\ model simulation. The second term is dependent upon both $E$ and $T$, and is largely driven by the \LBT\ model simulation. The $\theta$ function which scales the first term makes explicit the switching between them, at vituality \qswitch. See Ref.~\cite{JETSCAPE:2021ehl} for further discussion. 

%% file: BayesianInference.tex
\section{Bayesian Inference}
\label{sec:simulationsAndInferece}

Bayesian Inference is applied to constrain the following model parameters:

\begin{itemize}
\item  \alphas\ (denoted \alphasfix\ in  Eq.~\ref{eq:HTL-q-hat}), the coupling at the soft scale. Prior is uniform over $0.1 \leq \alphas \leq 0.5$.
\item \qswitch, the transition scale between the lower and higher virtuality stages of the simulation, that can be interpreted as the average value of $\qhat\tau$. Prior is uniform over $1~\mathrm{GeV}\leq{\qswitch}\leq10~\mathrm{GeV}$.
\item  \tstart, the start time of jet modification. Prior is uniform over $0\leq \tstart\leq$~1.5~fm/c.
\item Parameters $c_1 , c_2, c_3$ (Eq.~\ref{Eq:fmu}), which control the modification of \qhat\ with increasing virtuality. Priors are uniform in the logarithm of each parameter in the ranges $-5 \leq \log(c_{1,2}) \leq \log(10)$ and $-3 \leq \log(c_3) \leq \log(100)$. 
\end{itemize}

Prior distributions were determined by incorporating prior physics knowledge and approximate studies of the sensitivity of $\qhat$ to changes in $c_1$, $c_2$ and $c_3$.

A multi-dimensional normal distribution is used for the likelihood term. Following the procedure in Ref.~\cite{JETSCAPE:2021ehl}, a Markov Chain Monte Carlo procedure~\cite{Hogg:2017akh} is employed to explore the parameter space to determine the parameter posterior distributions. However, direct computation of all choices of parameter values with the precision required to perform Bayesian Inference is prohibitively expensive. Instead, simulations are performed at selected coordinates in parameter space (referred to as ``design points''), with a Gaussian Process Emulator (GPE) trained to interpolate between design points. The GPE serves as a computational fast surrogate model for calculations in parameter space.

\subsection{Simulations}
\label{sect:Simulations}

Simulations were carried out using \JETSCAPE\ v3.5\footnote{\JETSCAPE\ v3.5 was slightly modified to use \PYTHIA\ 8.235}, employing the physics modules described in Sec.~\ref{sec:physicsModel}. Each event propagates partons from a single hard scattering through a 2+1D calibrated medium~\cite{Bernhard:2019bmu}.
We utilized 20--40 pre-computed medium profiles per 1\% interval in centrality, randomly selecting a single  profile for each separate hard scattering event.

At each design point, simulations were performed for all collision energies of the datasets considered (Sect.~\ref{sec:experimentalDatasets}), taking into account the fiducial and kinematic acceptances of each measurement. Since the final--state hadrons of each event have been recorded, additional observables can be explored in future analyses. To optimize the utilization of computing resources, the number of simulated events was chosen so that the statistical precision of key calculated observables matched the magnitude of corresponding experimental uncertainties.

These simulations required $\mathcal{O}(5000-10000)$ core-hours per design point per collision energy on nodes with two AMD EPYC 7742 processors, which have 64 cores per processor.
Calculations are highly parallel and have significant IO requirements for loading pre-computed hydrodynamic profiles and storing all final-state hadrons.
The computations were run on three high-performance computing facilities~\cite{Towns2014,Expanse2022,Brown2021} and stored on the Open Storage Network~\cite{Kirkpatrick2021}.

An active--learning sampling scheme~\cite{cohn1996active} was used for strategic selection of design points for simulation. Active learning is a growing area in machine learning, which addresses the challenge of limited sample sizes by using the trained learning model to query subsequent sample points~\cite{mak2017information,chen2022adaptive,song2023ace}. In this analysis an active learning procedure sequentially selects batches of design points for simulation, guided by the trained Gaussian Process Emulator discussed below.

The active learning algorithm proceeds as follows. Given a total budget of 230 design points, we first select an initial batch of 40 design points using a Latin hypercube design~\cite{mckay2000comparison}, which provides a uniform coverage of the parameter space.
The simulated hadron and jet $\RAA{}$ at these initial design points are then used to train an initial GPE model. With this model in hand, the trained emulator is used to select a new batch of 20--40 design points for subsequent simulation. These points are optimized via a weighted sampling approach called importance support points (ISPs; \cite{huang:2022pqmc,mak:2018sp}), which targets two properties. First, such points mimic the probability distribution $[\sigma^2(\boldsymbol{\theta})]^l / \int [\sigma^2(\boldsymbol{\theta})]^l d \boldsymbol{\theta}$, where $\sigma^2(\boldsymbol{\theta})$ is the predictive variance of the GPE at parameters $\boldsymbol{\theta}$ over all observables. As such, new design points should target regions of the parameter space where $\sigma^2(\boldsymbol{\theta})$ is large, i.e., where the GPE is most uncertain. Second, given this distributional constraint, ISPs target design points that are well spaced-out, thereby exploring the parameter space; such a space-filling property allows for good predictive performance for GPEs \cite{mak2018minimax}. This captures the exploration-exploitation trade-off fundamental to reinforcement learning \cite{kearns2002near}; in our experiments, $l=10$ appears to provide a good trade-off. We then iterate the steps of GPE training, ISP optimization and event simulation until the desired run size is reached. The final 30 design points are taken from a separate Latin hypercube design as a validation set.
 
In summary, the simulations in this analysis required $\mathcal{O}(5.5)$ million CPU core-hours. Following the completion of the simulation campaign, including observable calculation, post-processing, and quality assurance checks, we then convert the simulations into a form suitable for Bayesian inference.
In order to focus on the primary features of the model space, to reduce sensitivity to statistical fluctuations, and to improve numerical efficiency, a Principal Component Analysis (PCA) is performed.
The number of features to retain was determined via a sensitivity study, with selection criteria requiring good description of the feature variance of the physical observables while reducing sensitivity to random fluctuations.
Each feature is modeled with a separate GPE.
The uncertainties in the prediction from the GPE are added to their experimental counterparts when comparing to data.

\subsection{Experimental datasets}
\label{sec:experimentalDatasets}

This analysis incorporates a broad set of inclusive hadron and jet \RAA\ measurements at RHIC and the LHC, for \AuAu{} collisions at $\snn=0.2\;\tev$ and \PbPb{} collisions at $\snn=2.76$ and $5.02\;\tev$. Due to the large-scale computational nature of the analysis, it was necessary to impose a cutoff date on the data considered. All such measurements which were published or submitted for publication prior to Feb. 2022 are included in the analysis, corresponding to 729 data points\footnote{The most recent ATLAS hadron \RAA\ measurement~\cite{ATLAS:2022kqu} was published after the cutoff date and is not considered in the analysis.}. In comparison, the previous Bayesian calibration of \qhat\ by \JETSCAPE~\cite{JETSCAPE:2021ehl} was based only on inclusive hadron \RAA\ data from a limited set of measurements~\cite{ PHENIX:2012jha,Aad:2015wga, Khachatryan:2016odn}, corresponding to 66 data points.

  \begin{center}
\begin{table*}[!tbh]
\centering
    \begin{tabular}{|c||c|c|c|c|c|}
    \hline
    \multicolumn{6}{|c|}{Inclusive hadron \RAA} \\
    \hline
    Collab./ref. & System; \sqrtsNN\  & Species & Accept.  & centr. & \pT\ range  \\
     & [TeV] &  &  & \% &  [GeV/c] \\
    \hline \hline
    STAR~\cite{STAR:2003fka} & \AuAu; 0.2 & charged & $|\eta|<0.5 $ & [0,40]  & [9,12] \\
    ALICE~\cite{Acharya:2018qsh}  & \PbPb; 2.76, 5.02 & charged & $|\eta|<0.8$ & [0,50] & [9,50]  \\
    ATLAS~\cite{Aad:2015wga} & \PbPb; 2.76 & charged & $|\eta|<2$  & [0,40] & [9,150]  \\
    CMS~\cite{CMS:2012aa} & \PbPb; 2.76 & charged & $|\eta|<1.0$ & [0,50]  & [9,100]  \\
    CMS~\cite{Khachatryan:2016odn} & \PbPb; 5.02 & charged & $|\eta|<1.0$ & [0,50]  & [9,400]  \\   
    PHENIX~\cite{Adare:2008qa} & \AuAu; 0.2 & $\pi^0$ & $|\eta|<0.35$ & [0,50]  & [9,20]  \\
    ALICE~\cite{Abelev:2014ypa, Acharya:2018yhg} & \PbPb; 2.76 & $\pi^0$ & $|\eta|<0.7$ & [0,50] & [9,20]  \\
    ALICE~\cite{Abelev:2014laa, Adam:2015kca} & \PbPb; 2.76 & $\pi^\pm$  & $|\eta|<0.8$  & [0,40] & [9,20]  \\
    ALICE~\cite{Acharya:2019yoi} & \PbPb; 5.02 & $\pi^\pm$  & $|\eta|<0.8$  & [0,50] & [9,20]  \\
    \hline
    \end{tabular}
    \caption{Datasets used in the analysis: inclusive hadron \RAA.}
    \label{tab:Datasets_HadronRAA}
\end{table*}

\begin{table*}[!tbh]
    \centering
    \begin{tabular}{|c||c|c|c|c|c|c|}
    \hline
    \multicolumn{7}{|c|}{Inclusive jet \RAA} \\
    \hline
    Collab./ref. & System; \sqrtsNN  & type & \rr & Accept.  & centr. & \pT\ range  \\
    & [TeV] &  &  &  & \% & [GeV/c] \\
    \hline \hline
    STAR~\cite{Adam:2020wen} & \AuAu; 0.2 & charged & [0.2,0.4] & $|\eta|<1-\rr$ & [0,10]  & [15,30] \\
    ALICE~\cite{Adam:2015ewa} & \PbPb; 2.76 & full & 0.2 & $|\eta|<0.5$ & [0,30]  & [30,100] \\
    ALICE~\cite{Acharya:2019jyg} & \PbPb; 5.02 & full & 0.2,0.4 & $|\eta|<0.5$ & [0,10]  & [40,140] \\
    ATLAS~\cite{Aad:2014bxa} & \PbPb; 2.76 & full & 0.4 & $|\eta|<2.1$ & [0,50]  & [32,500] \\
    ATLAS~\cite{Aaboud:2018twu} & \PbPb; 5.02 & full & 0.4 & $|\eta|<2.8$ & [0,50]  & [50,1000] \\    
    CMS~\cite{Khachatryan:2016jfl} & \PbPb; 2.76 & full & [0.2,0.4] & $|\eta|<2.0$ & [0,50]  & [70,300] \\
    CMS~\cite{Sirunyan:2021pcp} & \PbPb; 5.02 & full & [0.2,1.0] & $|\eta|<2.0$ & [0,50]  & [200,1000]\\
    \hline
    \end{tabular}
    \caption{Datasets used in the analysis: inclusive jet \RAA.}
    \label{tab:Datasets_JetRAA}
\end{table*}
 \end{center}

Tables~\ref{tab:Datasets_HadronRAA} and \ref{tab:Datasets_JetRAA} show the datasets used in the analysis, for inclusive hadron \RAA\ and inclusive jet \RAA\ respectively. To account for the range of model applicability, the analyzed centrality and kinematic ranges specified in the tables do not always correspond to the full range of the published data. This analysis only uses measurements in the centrality range  $0-50\%$, and inclusive hadron \RAA\ is used only in the range $\pT>9\;\gev$. These limitations will be relaxed in future analyses.

The analysis utilizes the statistical and systematic uncertainties specified in the experimental publications. While the uncertainty covariance matrix is required for Bayesian Inference, it is typically not provided in the experimental publications.
The uncertainty covariance matrix is therefore estimated where possible based on publicly available information, with separate treatment of statistical uncertainty and source-by-source systematic uncertainties.
If insufficient information is available for this estimate from the publication for any given source, an uncertainty correlation length is employed, as described in Ref.~\cite{JETSCAPE:2021ehl}.  The covariance for these sources in this case can be written as
\begin{align}
    \Sigma_{k,ij} = \sigma_{k,i} \sigma_{k,j} \exp\left[-\left|\dfrac{p_{k,i}-p_{k,j}}{\ell_k}\right|^2\right],
\end{align}
where $p_{k,i}$ is the $i^\text{th}$ \pT\ value of experimental measurement $k$ and $\ell_k$ is the covariance length, with a nominal value of 0.2.  The $p_{k,i}$ transverse momentum values are linearly rescaled so that all values from the given measurement lie within $[0,1]$. The systematic uncertainties of different data sets are assumed to be independent, with the exception of uncertainties in the luminosity and nuclear thickness calculations for inclusive jet and leading hadron measurements by a single experiment at a given collision energy.  However, these uncertainties are negligible compared to others, and for this reason they are not treated differently. The covariance matrices constructed from different sources are then added together to form the total covariance matrix.

%% file: Results.tex
\section{Results}
\label{sec:results}

\begin{figure}[tbhp!]
\begin{center}
\includegraphics[width = 0.45\textwidth]{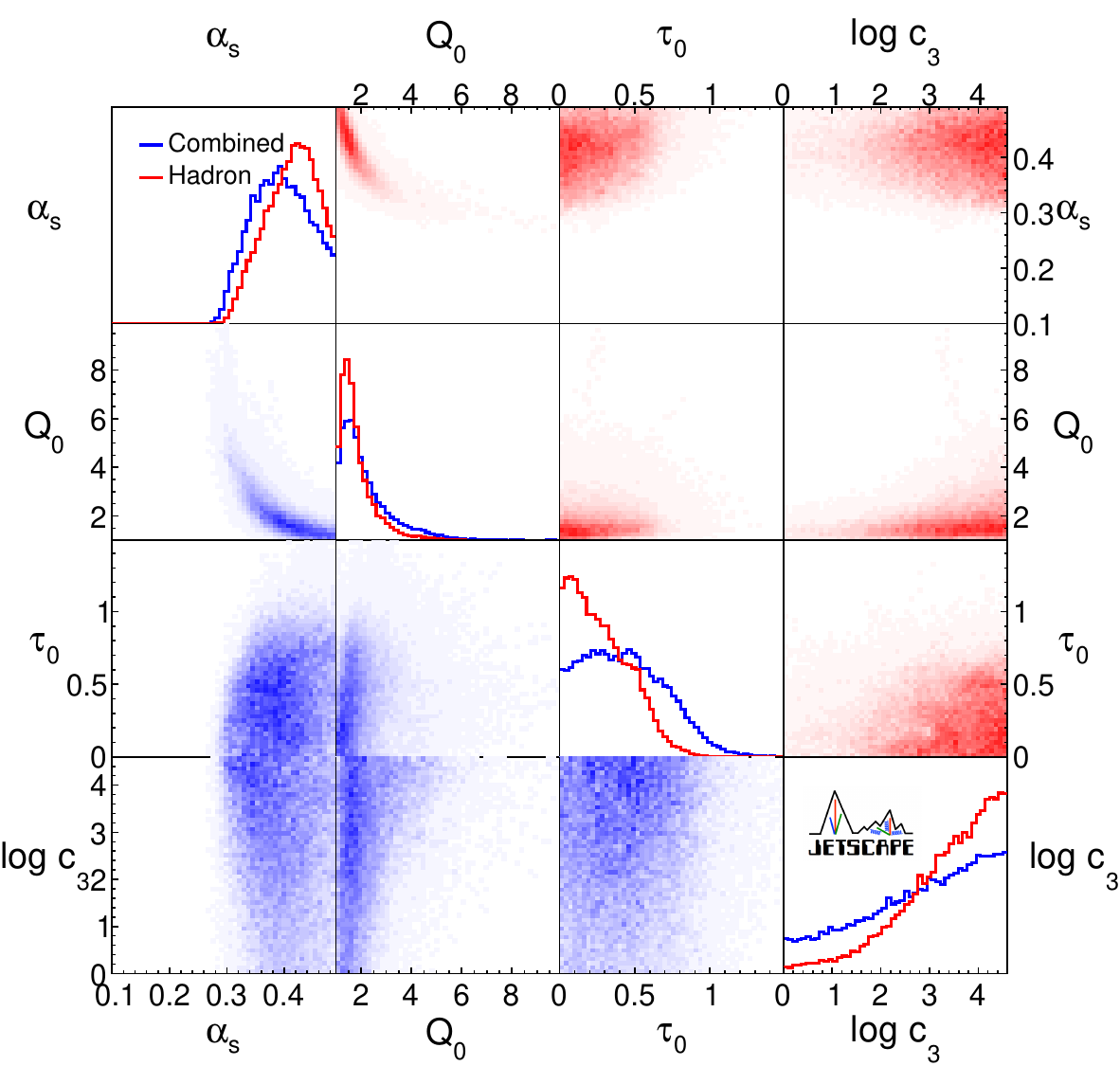}
\end{center}
\caption{Parameter posterior distributions (diagonal) and parameter--pair correlations (off-diagonal), determined by the combined analysis of inclusive hadron and jet \RAA\ data (blue), or inclusive hadron \RAA\ only (red). Parameters $c_1$ and $c_2$ are not constrained significantly by the calibration and their distributions are not shown here; Fig.~\ref{fig:PosteriorParameterAlt} shows the full set of parameter posterior distributions and correlations.}
\label{fig:PosteriorParameter}
\end{figure}

\begin{figure*}[tbhp!]
\begin{center}
\includegraphics[width = 0.035\textwidth]{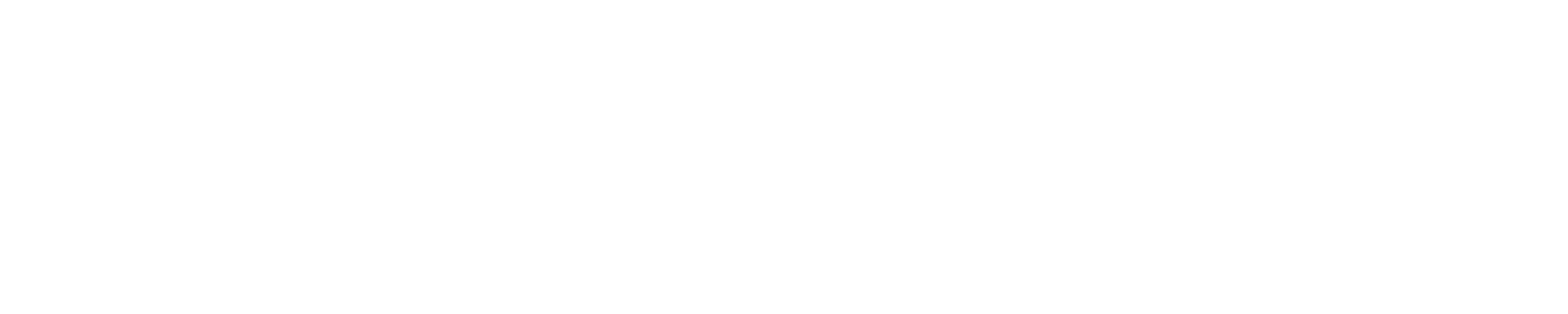}
\includegraphics[width = 0.45\textwidth]{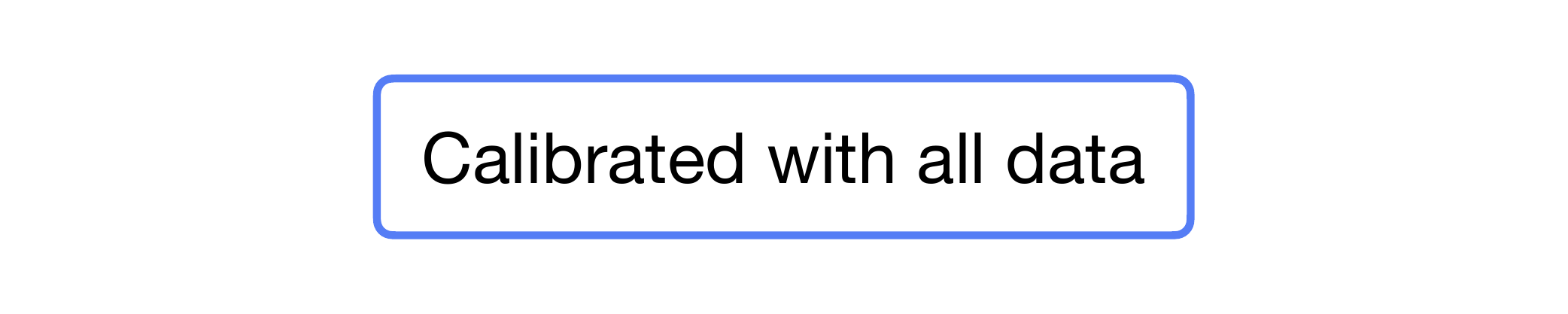}
\includegraphics[width = 0.45\textwidth]{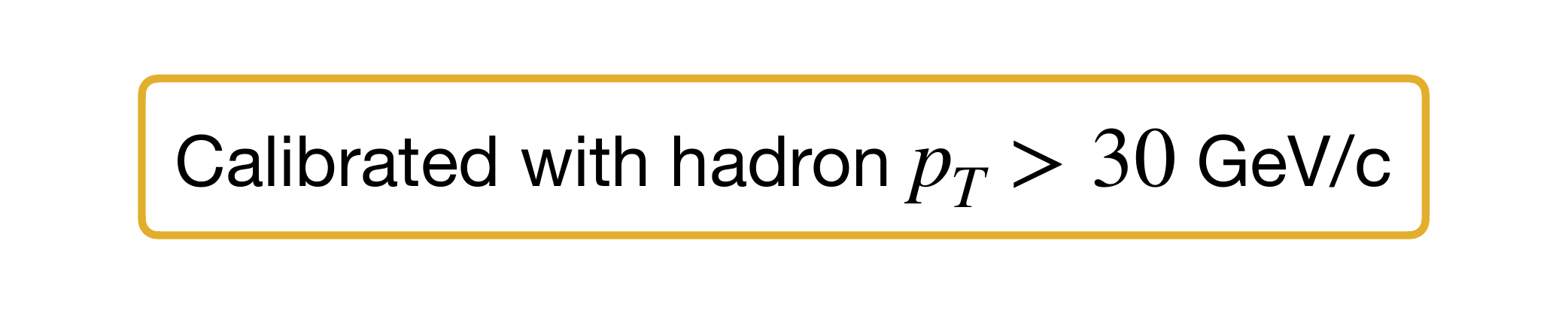}
\includegraphics[width = 0.45\textwidth]{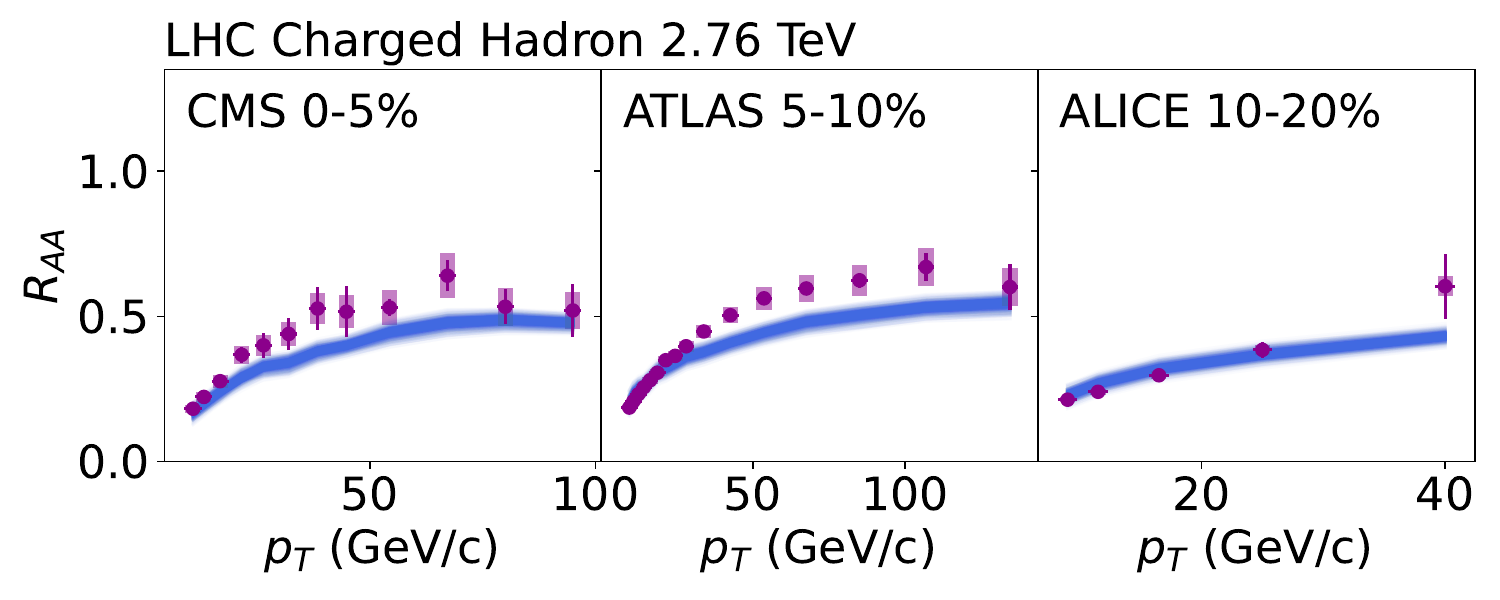}
\includegraphics[width = 0.45\textwidth]{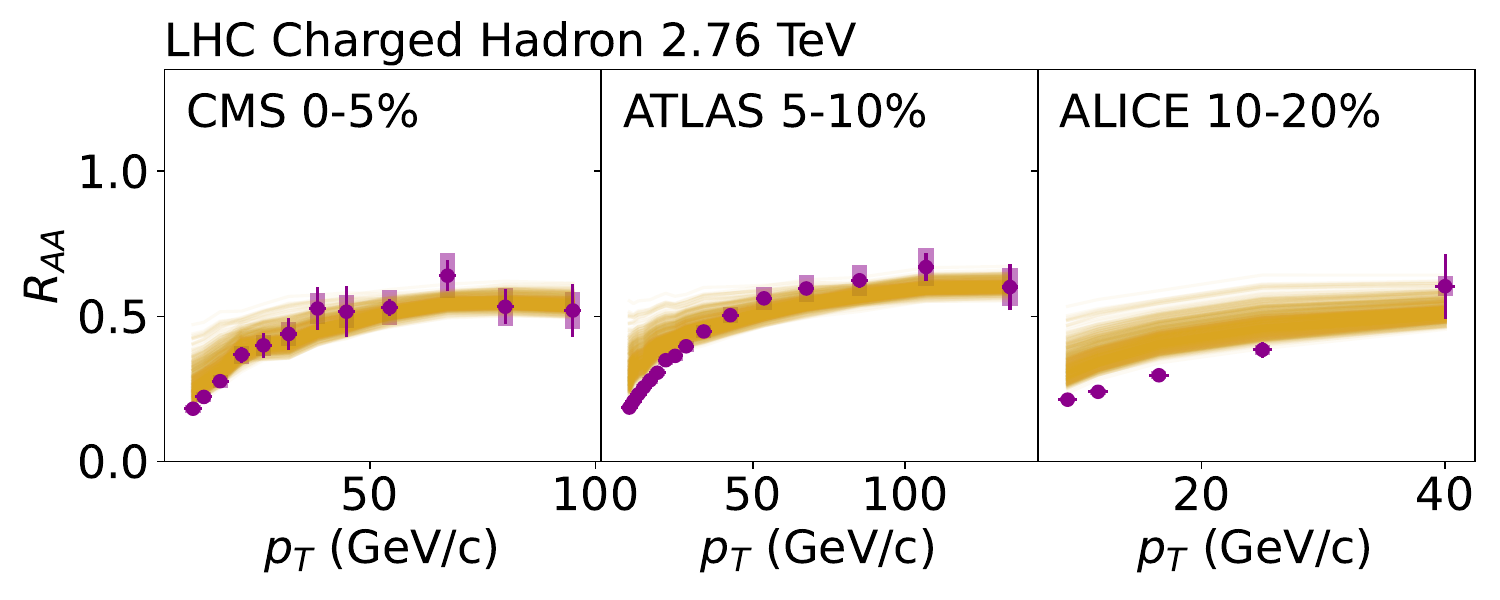}
\includegraphics[width = 0.45\textwidth]{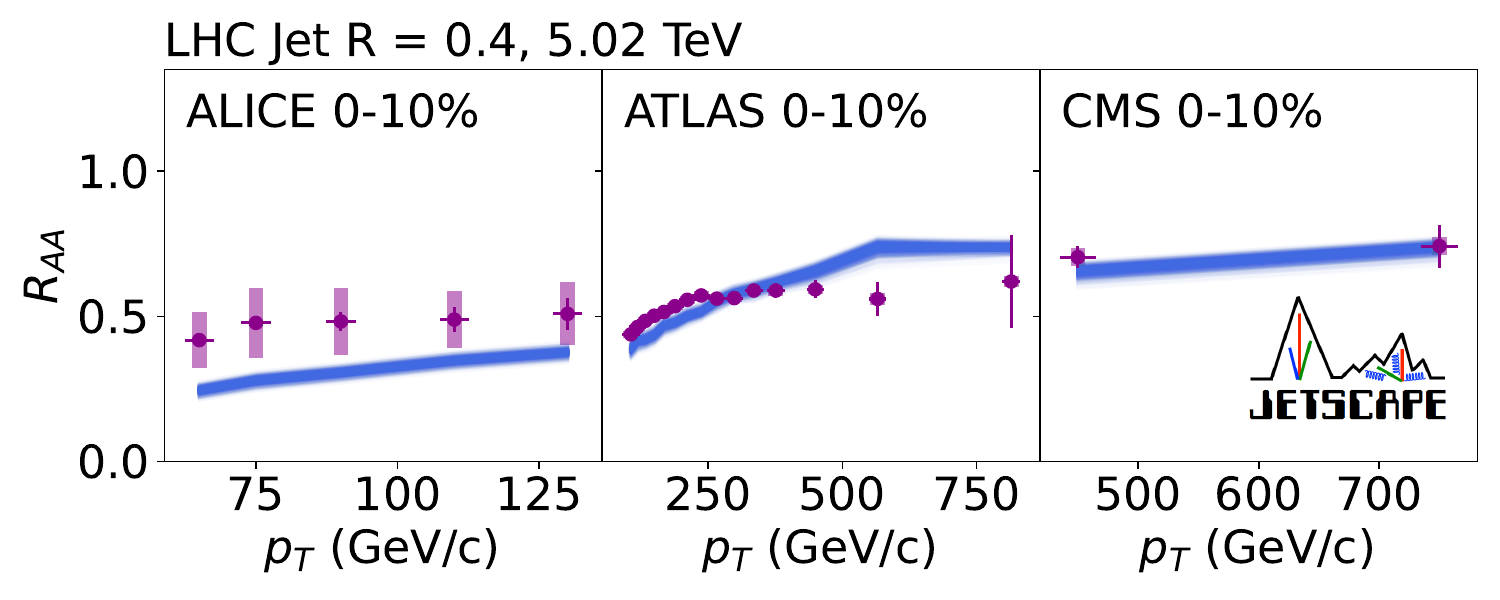}
\includegraphics[width = 0.45\textwidth]{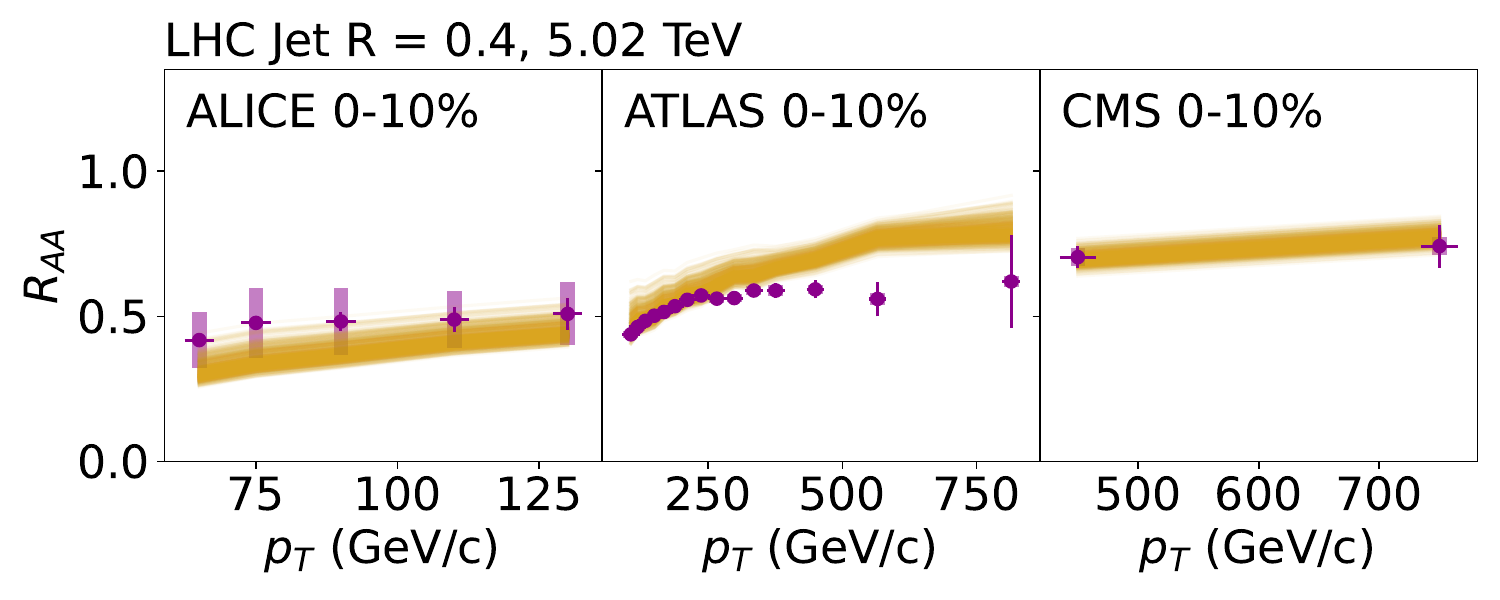}
\includegraphics[width = 0.45\textwidth]{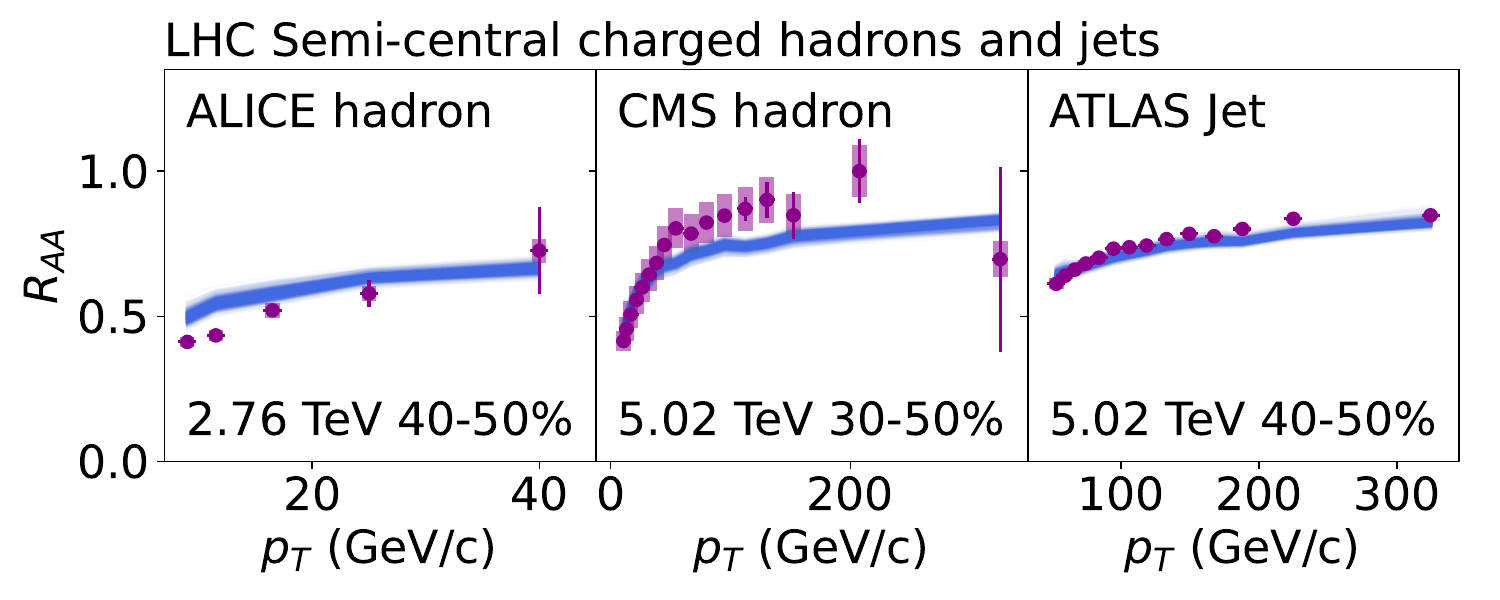}
\includegraphics[width = 0.45\textwidth]{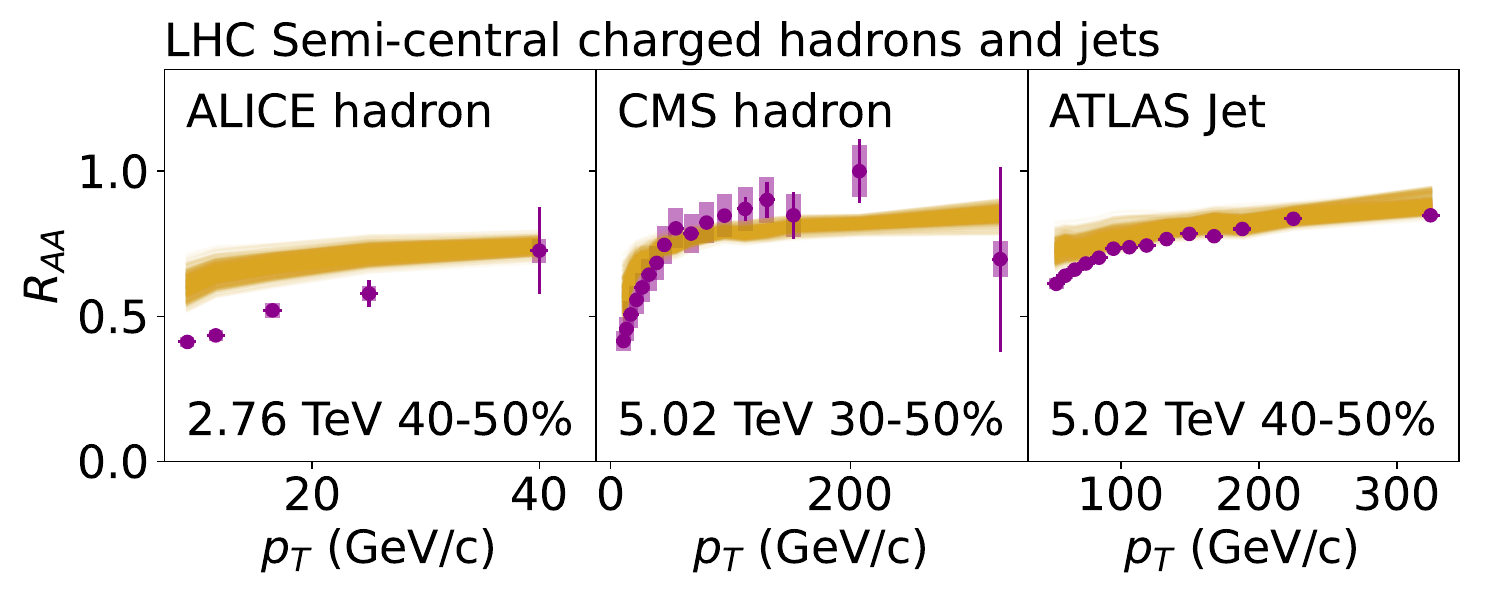}
\includegraphics[width = 0.32\textwidth]{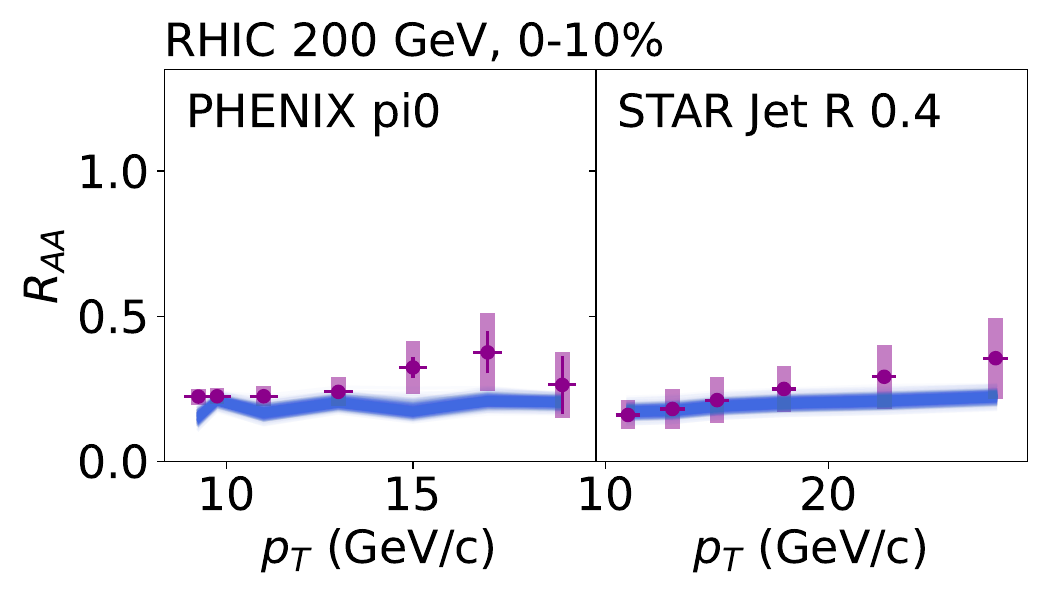}
\includegraphics[width = 0.32\textwidth]{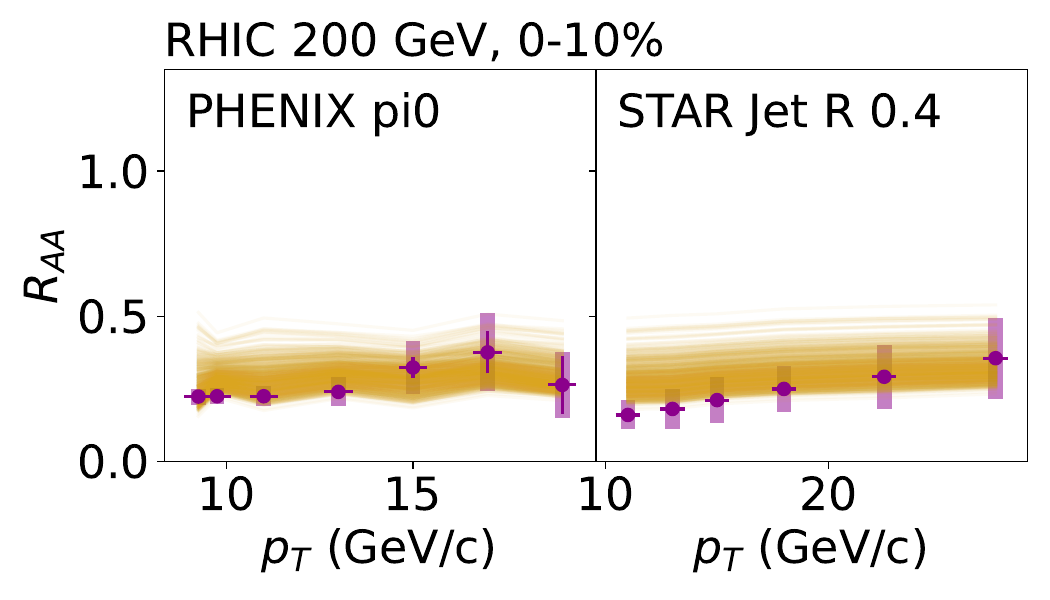}
\end{center}
\caption{Selected measurements of hadron and jet \RAA\ at $\sqrtsNN= 0.2$, 2.76, and 5.02 TeV at various centralities, compared to posterior predictive distributions for the combined calibration (left set of panels; blue posterior distributions) and the calibration based solely on hadrons with $\pT>30$ \GeVc\ (right set of panels; orange posterior distributions). The same data are shown in both sets of panels (purple).}
\label{fig:ExamplePosterior}
\end{figure*}

Initial comparison of calculations utilizing a virtuality--dependent formulation of \qhat\ (Sect.~\ref{sect:qhat}) with a limited subset of data, and without comprehensive parameter assessment based on Bayesian Inference, is provided in Ref.~\cite{JETSCAPE:2022jer}. In this section, the multi-observable nature and kinematic range of the data are utilized to explore systematic sensitivity of posterior distributions to choice of observable and phase--space coverage. If \qhat\ is a universal property of the QGP and the employed theoretical framework is accurate, then the posterior distributions should be consistent for variations in choice of observable and phase space coverage. An observation of tension in posterior distributions with such variations could arise from limitations in the theoretical formulation of \qhat, including its dependence on $E$ and $T$, and limitations in the QGP model that is used in the bulk simulations. Tension may also indicate that interpretation of \qhat\ as an intensive property of the medium is not strictly correct.

Posterior distributions of \qhat\ are presented in terms of \qhatTcubed. Differential studies of posterior distributions are presented for several choices of observable, phase space coverage, and event centrality. In order to compare calibrations, the posterior distribution of \qhatTcubed\ is shown for a reference quark energy $\Eref=100$ GeV, either as a function of $T$ or at a reference temperature $\Tref=200$ MeV.

\subsection{Posterior distributions}
\label{sec:posteriors}

The model parameters which are calibrated by Bayesian Inference are presented in Sect.~\ref{sec:simulationsAndInferece}. The baseline calibration, denoted ``Combined,'' utilizes the full set of inclusive hadron and jet \RAA\ data in Tabs.~\ref{tab:Datasets_HadronRAA} and \ref{tab:Datasets_JetRAA}. Fig.~\ref{fig:PosteriorParameter}, blue distributions, shows parameter posterior distributions and parameter pair correlations from the Combined calibration. Fig.~\ref{fig:ExamplePosterior}, left set of panels, shows selected hadron and jet \RAA\ measurements compared to posterior predictive distributions from the Combined calibration. Sect.~\ref{app:FullPosteriorPredict} presents the full set of jet and hadron \RAA, together with posterior predictive distributions from the Combined calibration.

These figures also show results from two of the alternative calibrations, chosen to illustrate specific points in the following discussion. Fig.~\ref{fig:PosteriorParameter}, red distributions, show the parameter posterior distributions and parameter pair correlations with a calibration based solely on inclusive hadron \RAA\ data but without further selection. Fig.~\ref{fig:ExamplePosterior}, right set of panels, compares the same data as in the left panels, in this case with a calibration based solely on hadron \RAA\ data in the range $\pT>30$ \GeVc.

When comparing the posterior distributions in the left and right sets of panels in Fig.~\ref{fig:ExamplePosterior} it is important to note that low-\pT\ hadron \RAA\ measurements have the highest relative precision of all the measurements considered, and therefore provide strong constraints on the Combined-data posterior. This precision difference is also reflected in the broader Credible Interval (CI, 90\%) for the high-\pT\ hadron calibration. 

\subsection{Combined analysis of inclusive hadron and jet \texorpdfstring{\RAA{}}{RAA}}
\label{sec:CombinedAnalysis}

The parameter posterior distributions from the combined analysis (Fig.~\ref{fig:PosteriorParameter}, blue) exhibit the following features:

\begin{itemize}

\item The coupling parameter \alphas\ is constrained to the range 0.3-0.5, with an approximately symmetric distribution that peaks near 0.4. 

\item The  model switching parameter \qswitch\ lies predominately in the range 1-2 GeV, with a tail extending to 4-6 GeV. A similar posterior distribution for \qswitch\ was observed in the previous \JETSCAPE\ Bayesian calibration of \qhat\ based solely on inclusive hadron \RAA~\cite{JETSCAPE:2021ehl}. 

\item A mild constraint on the posterior of \tstart\ is observed, with preferred values below 1~fm/$c$. 

\item Anti-correlation of \alphas\ and \qswitch\ is observed, which is characteristic of the multi-stage approach of this model: a larger value of \qswitch\ corresponds to an increase in the time spent in the \LBT\ stage of the jet evolution, in which more gluon radiation is emitted than in the \MATTER\ stage for the same value of coupling strength. A lower value of \alphas\ is therefore needed to describe the data correctly. 

\item A preference for larger $c_3$ values is observed. A peak is not observed within the range of priors of the calibration, indicating that a preferred value of $c_3$ may lie at large values. However, addressing this point will require a new calibration with a larger range of prior for $c_3$, and is beyond the scope of the present work.

\item The values of $c_1$ and $c_2$ are not constrained significantly, and for clarity are not shown in Fig.~\ref{fig:PosteriorParameter} (see Fig.~\ref{fig:PosteriorParameterAlt} for the full set of parameter posterior distributions and correlations). Future study will explore this lack of constraint on $c_1$ and $c_2$, which can arise from data sensitivity, theoretical limitations, or other causes. 

\end{itemize}

\begin{figure}[tbhp!]
\begin{center}
\includegraphics[width = 0.45\textwidth]{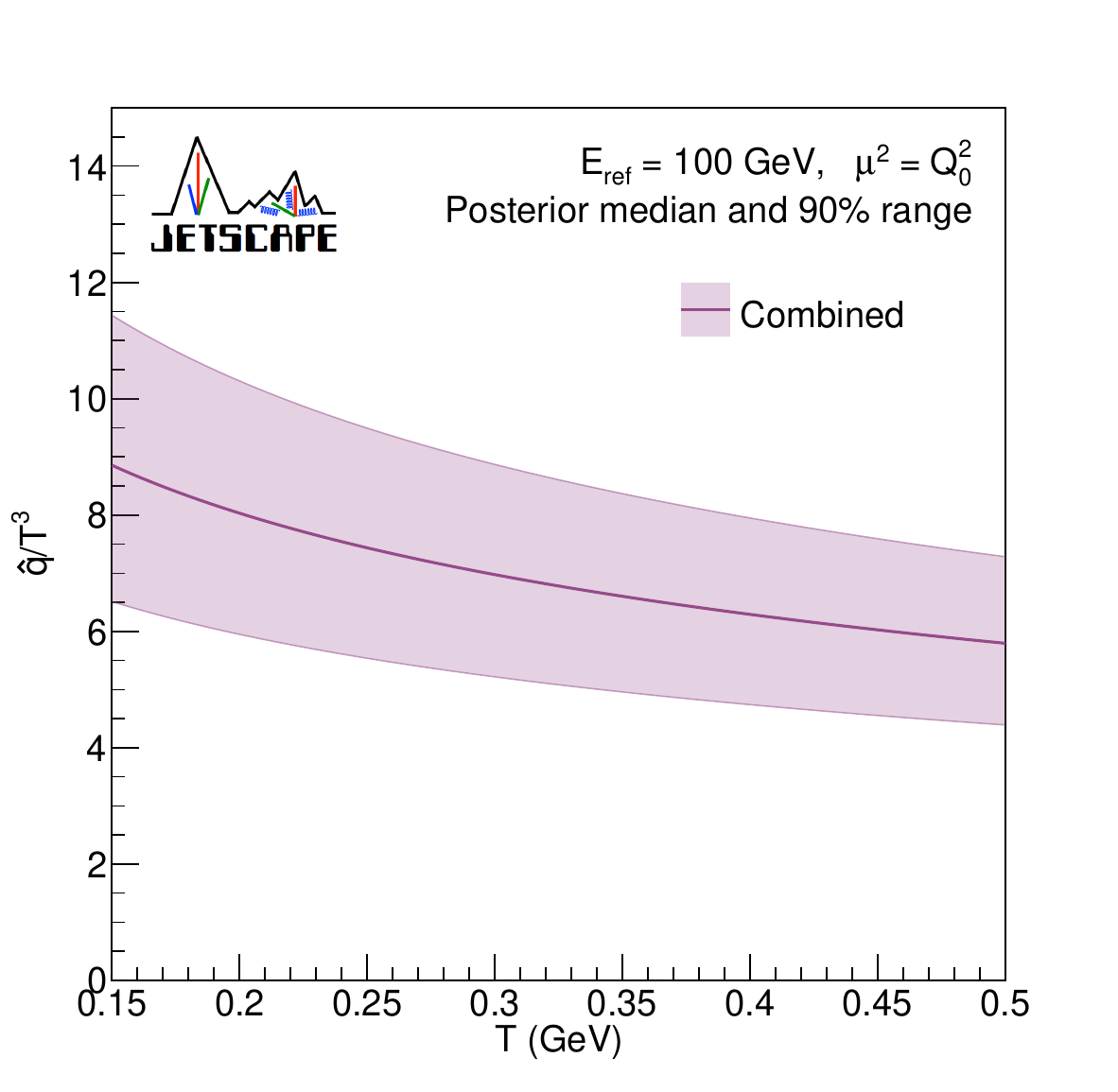}
\end{center}
\caption{Posterior distribution of \qhatTcubed(Eq.~\ref{Eq:fmu}) as a function of $T$, for a quark with energy $\Eref=100$ GeV and $\mu^2=\qswitch^2$, derived from the parameter posterior distribution of the Combined analysis. Lines show median values and bands show 90\% credible intervals.}
\label{fig:QHat}
\end{figure}

Figure~\ref{fig:QHat} shows the Combined--analysis posterior distribution of \qhatTcubed\ for a quark at $\Eref=100$ GeV as a function of $T$. The \qhatTcubed\ distribution is shown for $\mu^2=\qswitch^2$, whose distribution is determined by the Bayesian analysis,
and Eq.~\ref{Eq:fmu} reduces to Eq.~\ref{eq:HTL-q-hat}. The value of \qhatTcubed\ increases with decreasing $T$. This $T$-dependence is driven by both the measured data and the underlying physical model (Sect.~\ref{sect:qhat}). 

In the following sections we explore the differential dependence of \qhatTcubed\ on the observable and kinematic range used in the calibration.

\subsection{Comparison of inclusive hadron and inclusive jet calibrations}

\begin{figure*}[tbhp!]
\begin{center}
\includegraphics[width = 0.49\textwidth]{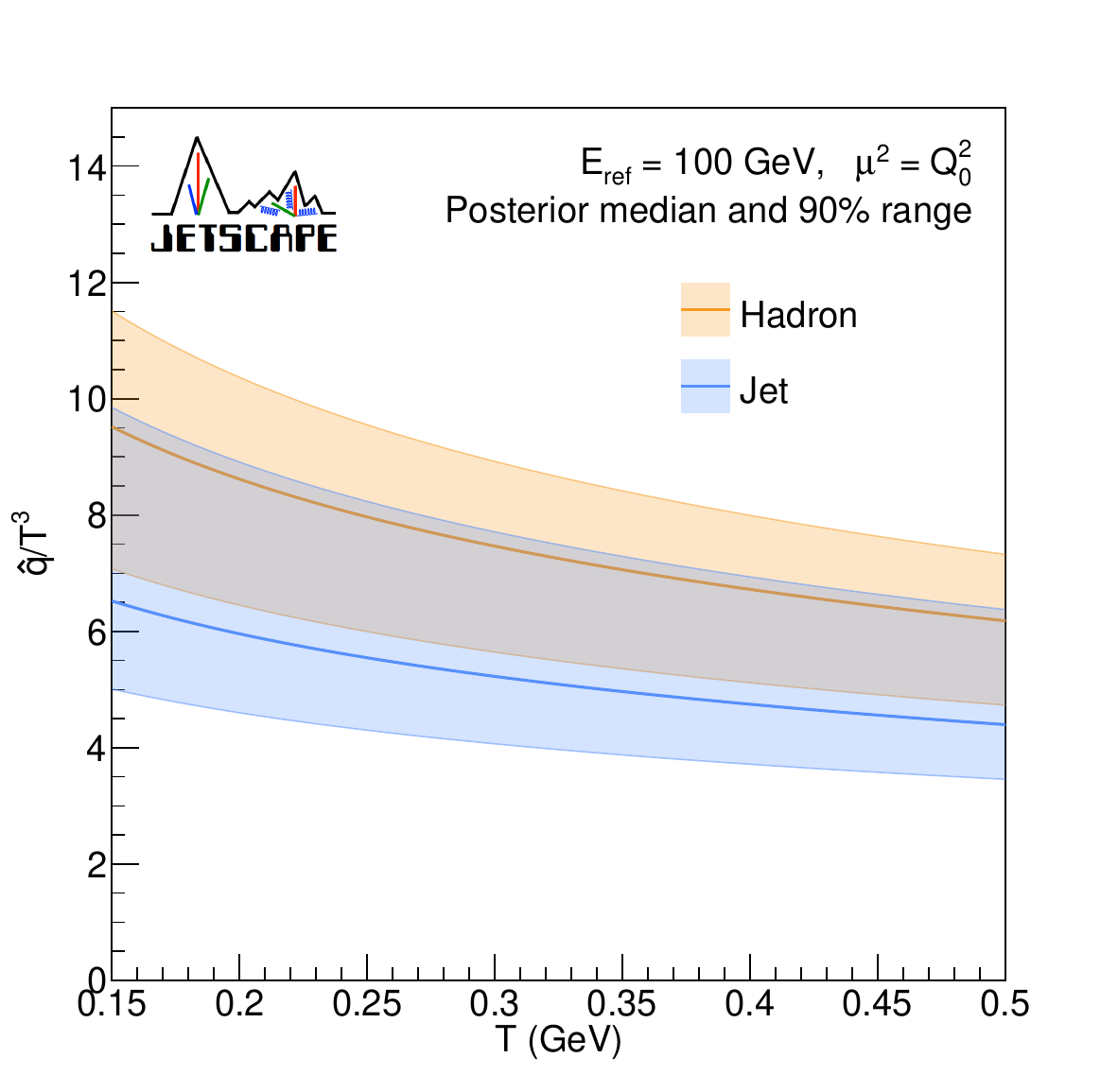}
\includegraphics[width = 0.49\textwidth]{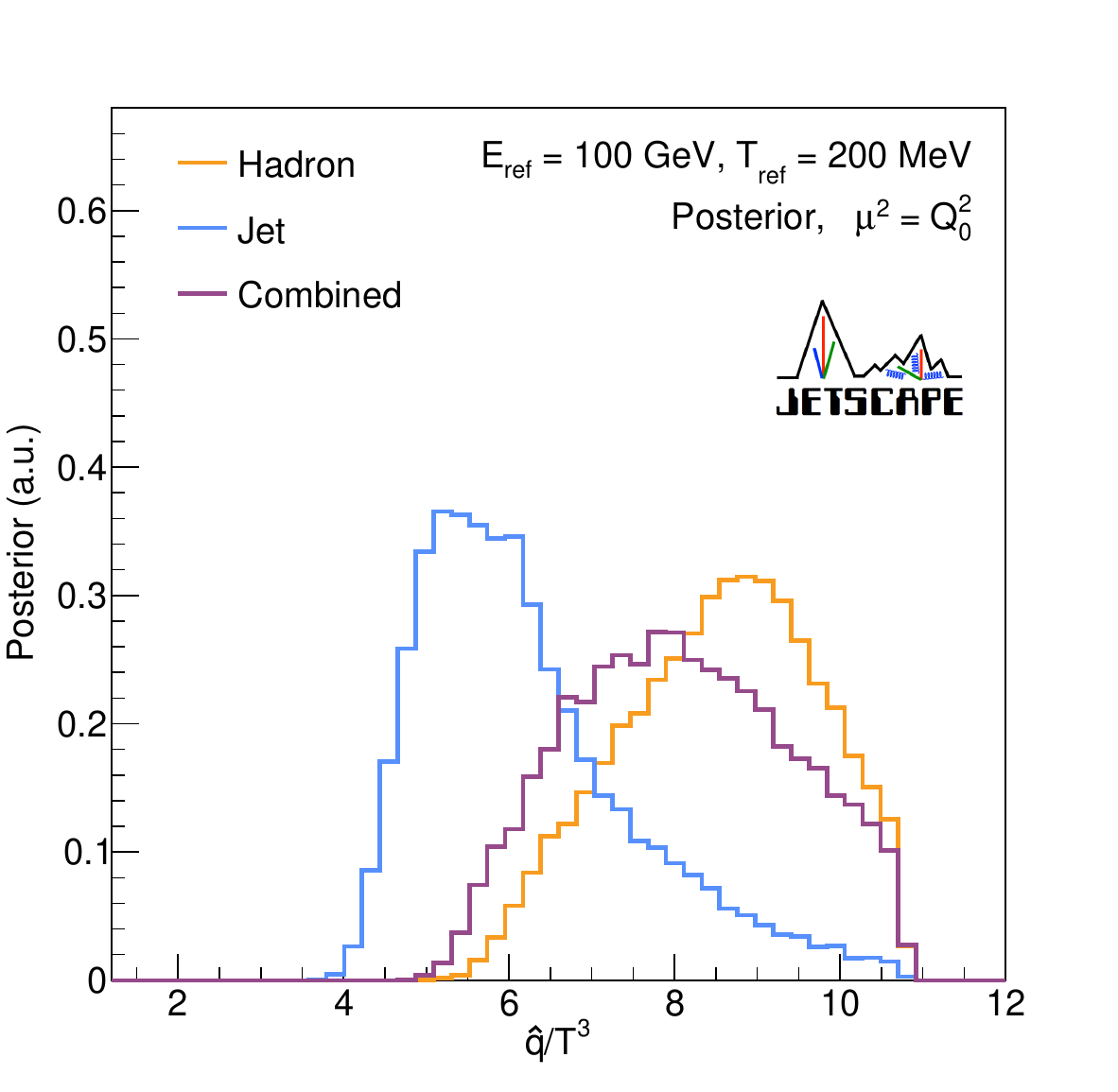}
\end{center}
\caption{Left: Posterior distributions (90\% CI) of \qhatTcubed\ for a quark with energy \Eref\ as a function of $T$, calibrated using hadron--only or jet--only \RAA\ data. Right: slice of hadron--only or jet--only calibrations in left panel at \Tref=200 MeV, compared to that for the Combined calibration. All distributions are normalized to unit integral.}
\label{fig:QHatJetHadron}
\end{figure*}

In order to explore the dependence of the posterior distributions on input data, Fig.~\ref{fig:QHatJetHadron} shows posterior distributions of \qhatTcubed\ for a quark with energy $\Eref=100$ GeV, calibrated separately on hadron or jet \RAA. The left panel shows the posterior distribution CI (90\%) as a function of $T$. While the hadron-only and jet-only posterior distributions are consistent and have similar shape within the CIs, the jet-only distribution brackets lower \qhatTcubed\ values. 

Figure~\ref{fig:QHatJetHadron}, right panel, shows the posterior distributions of \qhatTcubed\ for a quark with energy \Eref\ at temperature $\Tref=200$ MeV. The Combined posterior distribution is also shown. The most probable values for jet-only and hadron-only are markedly different, although the distributions overlap in a significant range, with an overlap fraction of 35.3\%. The Combined distribution lies between the two more-differential distributions, though with greater overlap with the hadron-only distribution.

Figure~\ref{fig:PosteriorParameter} elucidates the origin of this difference. The high--\pT\ hadron--only parameter posterior distributions in that figure (red) are qualitatively similar to the Combined case but are systematically narrower, preferring larger \alphas{}, smaller \qswitch{}, and smaller \tstart. The values of $c_{1}$ and $c_{2}$ remain unconstrained, while a larger value of $c_{3}$ is preferred. This bias towards stronger quenching generates larger values of \qhatTcubed, as shown in Fig.~\ref{fig:QHatJetHadron}.

Comparison of the hadron--only calibration in Fig.~\ref{fig:QHatJetHadron} to the previous JETSCAPE hadron--only calibration~\cite{JETSCAPE:2021ehl} is discussed in Sect.~\ref{sect:Compareqhatold}.

\subsection{Hadron kinematic selection}
\label{sec:hadronpTselect}

\begin{figure}[tbhp!]
\begin{center}
\includegraphics[width = 0.49\textwidth]{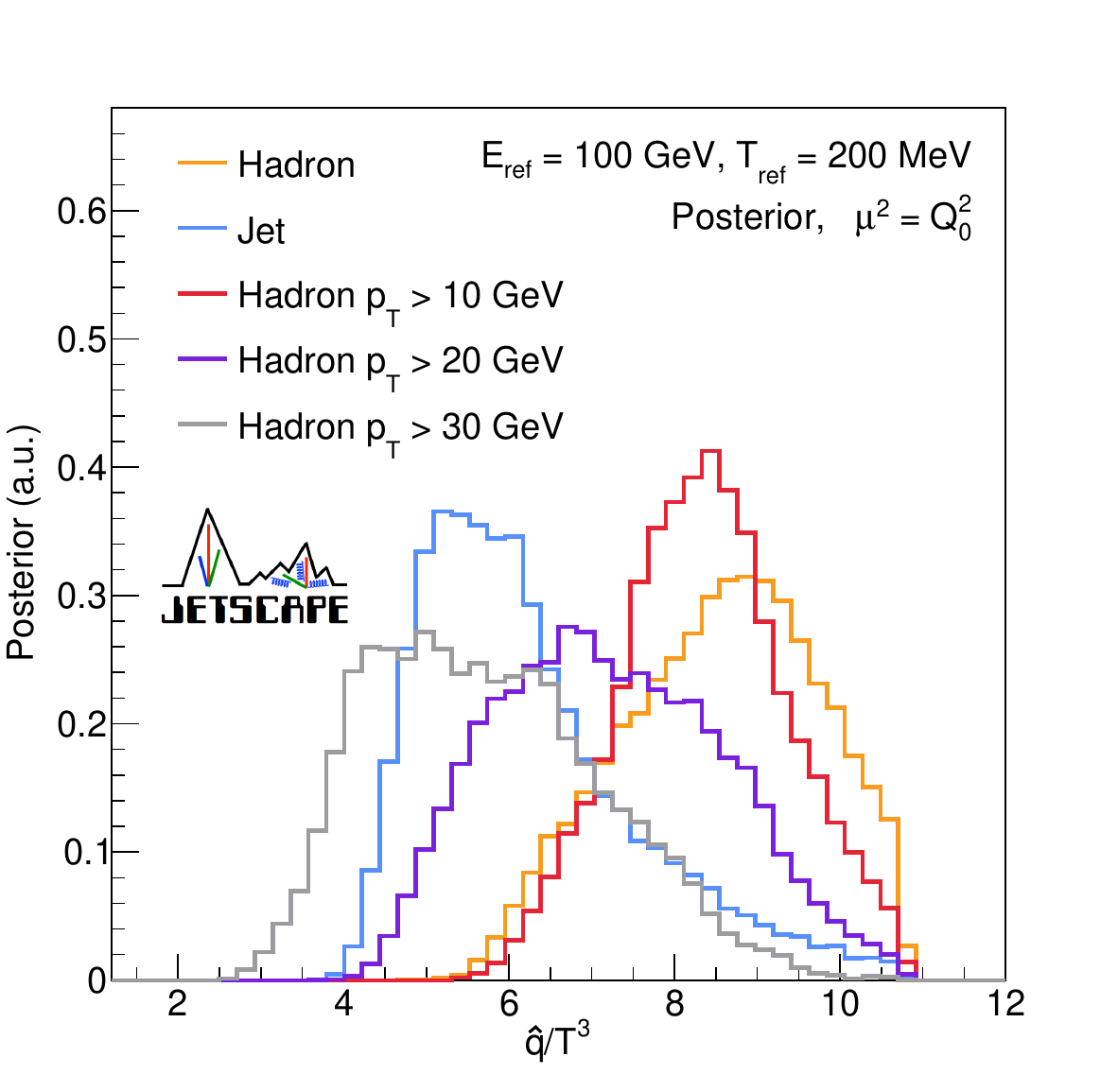}
\end{center}
\caption{Posterior distribution of \qhatTcubed\ for a quark with energy \Eref\ at temperature \Tref, calibrated using different input datasets. Distributions for jet--only (blue) and hadron--only (orange) are the same as in Fig.~\ref{fig:QHatJetHadron}. Distributions are also shown for hadron--only calibrations  with hadrons having $\pT>10$ (red), $\pT>20$ (purple), and $\pT>30$ (grey) \GeVc. All distributions are normalized to unit integral. 
}
\label{fig:QHatHadronPT}
\end{figure}

Figure~\ref{fig:QHatHadronPT} shows the dependence of the \qhatTcubed\ posterior distributions on hadron \pT\ range used in the calibration. The figure shows the jet--only and hadron--only distributions in Fig.~\ref{fig:QHatJetHadron}, together with hadron--only calibrations for hadrons with $\pT>10$, 20, or 30 \GeVc. These \pT-selected hadron--only distributions interpolate between the \pT-integrated hadron--only and jet--only calibrations, with hadron $\pT>10$ \GeVc\ most consistent with the hadron--only calibration, and hadron $\pT>30$ \GeVc\ most compatible with the jet--only calibration.

Inclusive hadron distributions are dominated by leading jet fragments, due to the combined effect of the falling inclusive jet spectrum with rising \pT\ and the falling jet fragmentation function with rising momentum fraction $z$. A PYTHIA 8 calculation for \pp\ collisions at $\sqrts=5.02$ TeV shows that the mean momentum fraction carried by the leading hadron in a jet is $\langle{z}\rangle\approx0.5$ over a broad range in \pTjet. The jet measurements in this analysis which have highest relative systematic precision cover the range $\pTjet \gtrsim 50$ \GeVc, corresponding to hadrons with $\pT \gtrsim 25$ \GeVc (for this illustration we neglect jet quenching effects on leading-hadron $\langle{z}\rangle$). The approximate agreement of posterior distributions of the hadron-only calibration with  $\pT>30$ \GeVc\ and the jet--only calibration, which probe jet quenching in a similar range of partonic kinematics, supports a picture in which \qhatTcubed\ is indeed independent of the way it is probed and may be a universal property of the QGP. However, their inconsistency with posterior distributions from lower-\pT\ probes indicates that the current model description of the partonic energy dependence of jet quenching may not be complete.

\subsection{High-\texorpdfstring{\pT{}}{pT} hadrons and jets}
\label{sec:HighpTJets}

\begin{figure}[tbhp!]
    \centering
    \includegraphics[width=0.49\textwidth]{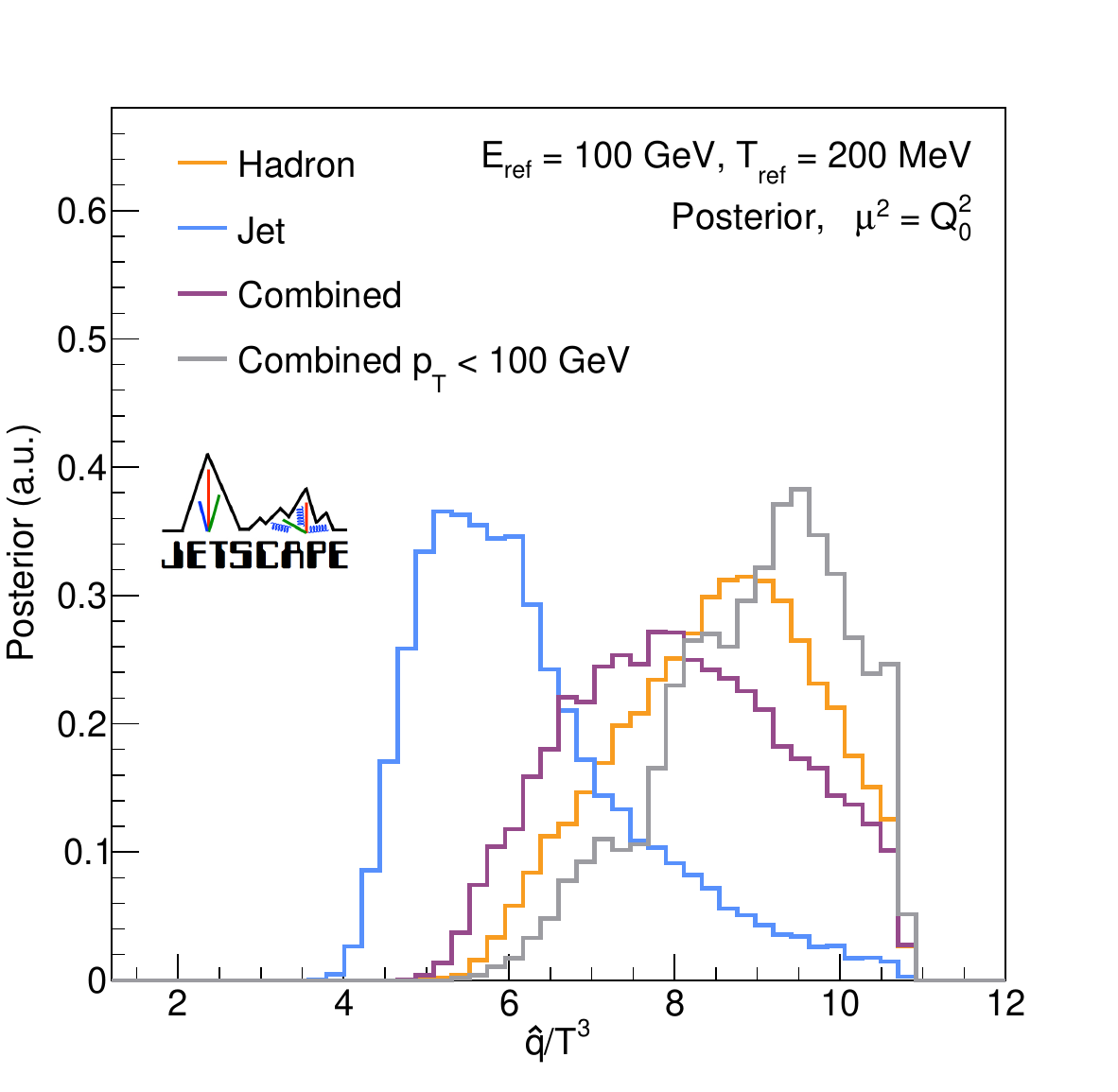}
    \caption{Posterior distribution of \qhatTcubed\ for a quark with $E=100$ GeV at $T=200$ GeV, calibrated using input data excluding $\pT{} > 100$ \GeVc{}. The distribution (grey) is compared to the Combined (purple), jet-only (blue) and hadron-only (orange) calibrations as shown in Fig.~\ref{fig:QHatJetHadron}. Distributions are normalized to unit integral.}
    \label{fig:QHatNoHighPt}
\end{figure}

There is tension between different measurements of inclusive jet \RAA\ at the highest \pT\ range used in this analysis (Tab.~\ref{tab:Datasets_JetRAA}). To assess the effect of such high-\pT\ measurements on the posterior distributions of \qhatTcubed, we carry out additional calibrations in which they are selectively excluded.

Figure~\ref{fig:QHatNoHighPt} shows the posterior distribution of a calibration excluding jet and hadron \RAA\ for $\pT>100$ \gev\ (grey), compared to the Combined, jet--only and hadron--only calibration posterior distributions shown in Fig.~\ref{fig:QHatJetHadron}, right panel. The \pT-restricted distribution is qualitatively similar to the hadron--only calibration, with a small shift towards larger $\qhatTcubed$.
Although the high--precision low-\pT\ hadron data dominate the calibration, this comparison shows that the relatively lower precision high-\pT\ hadron and jet data nevertheless have significant influence, shifting \qhat\ to lower values.

\subsection{Collision--centrality dependence}
\label{sec:centrality}

\begin{figure*}[tbhp!]
    \centering
    \includegraphics[width=0.325\textwidth]{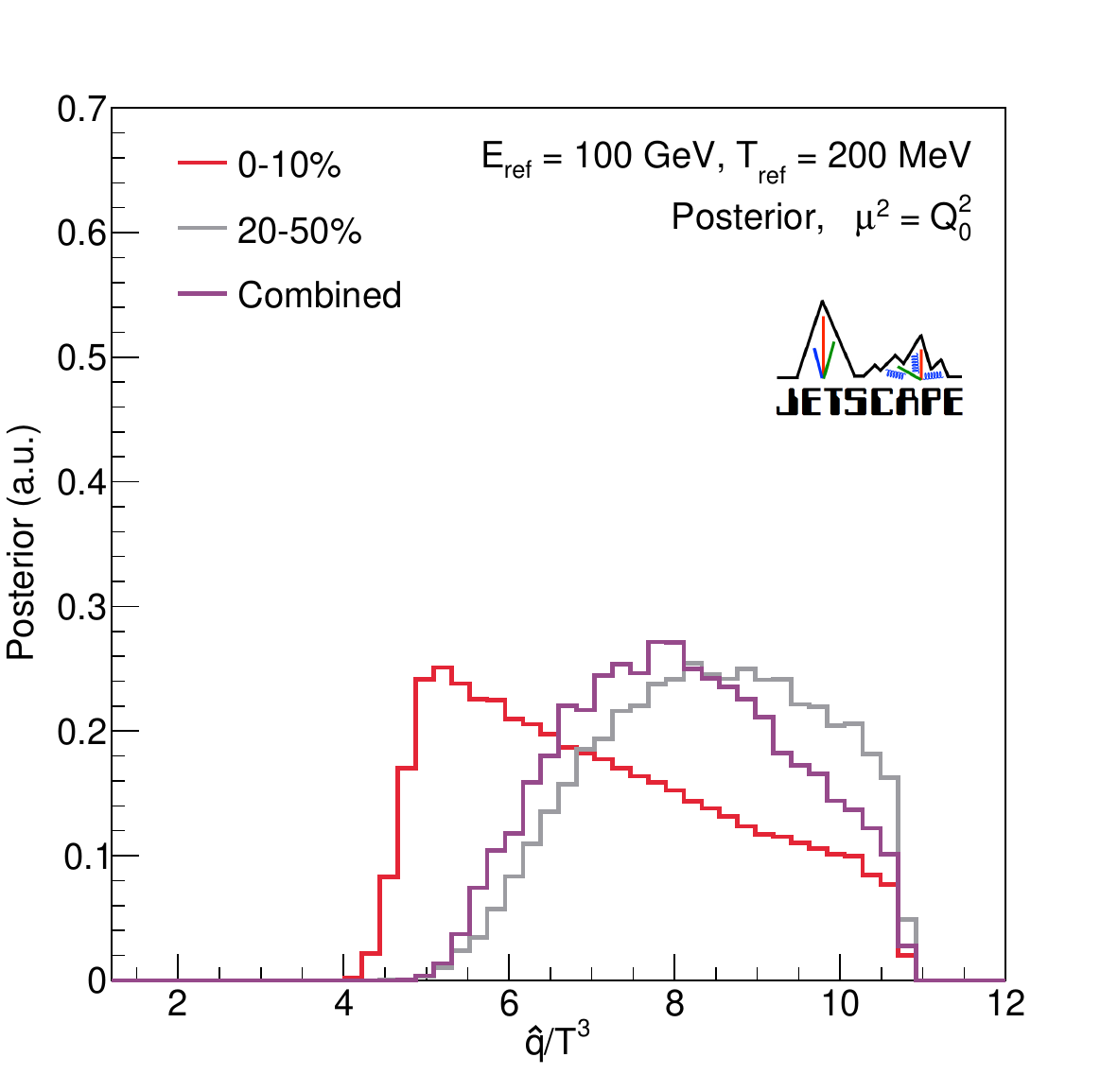}
    \includegraphics[width=0.325\textwidth]{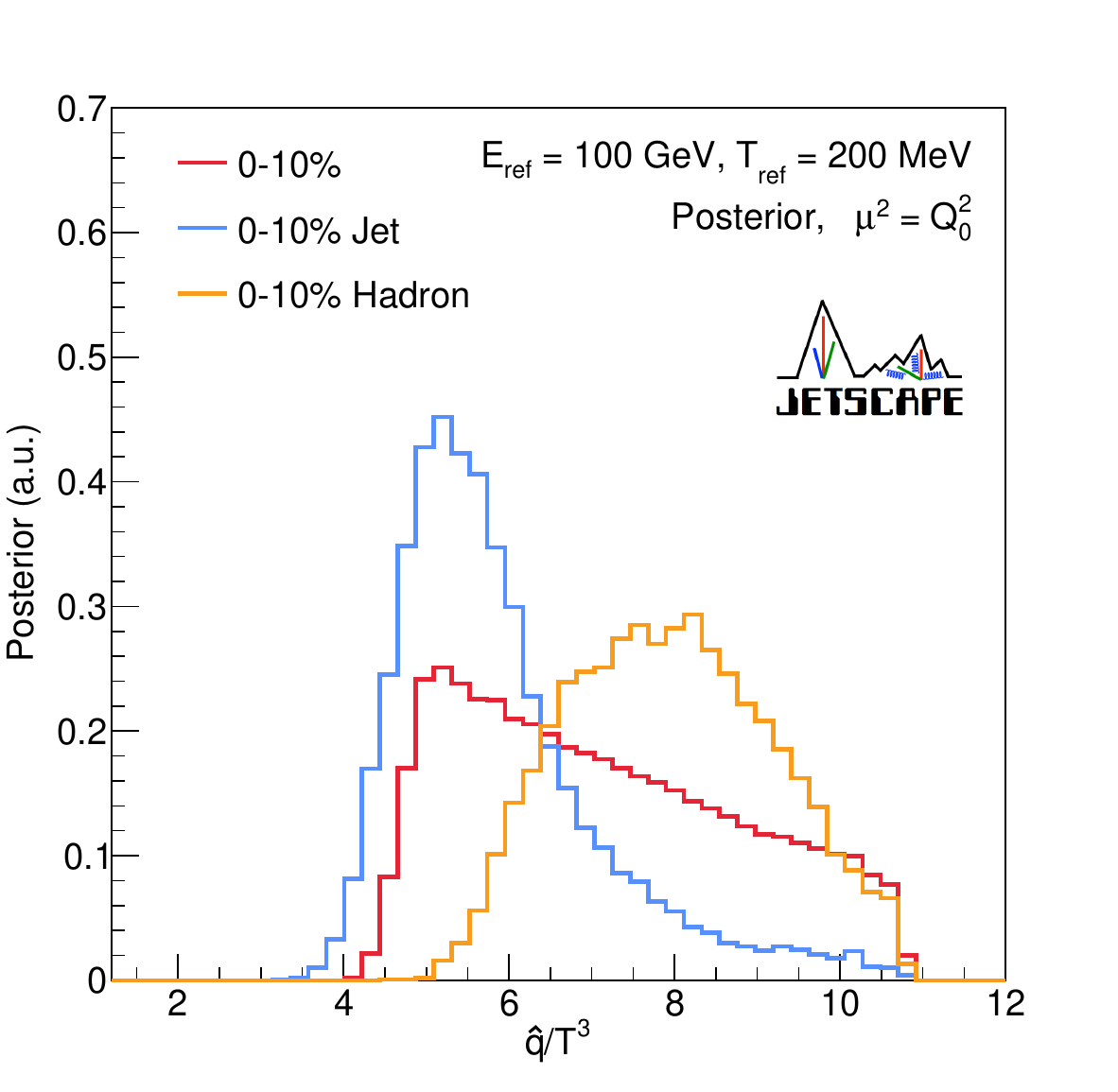}
    \includegraphics[width=0.325\textwidth]{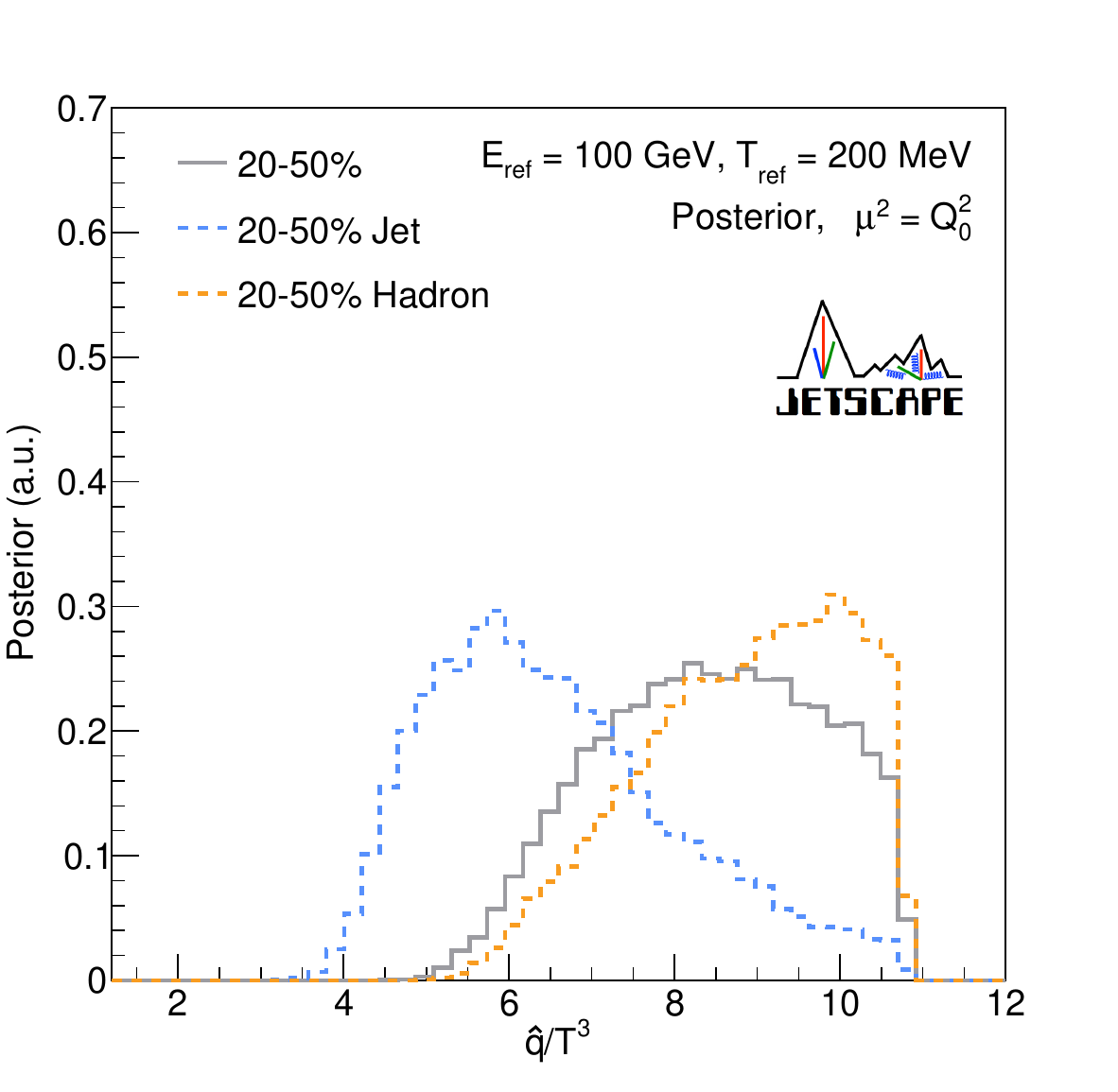}
\caption{Posterior distributions of \qhatTcubed\ for a quark with energy \Eref\ at temperature \Tref, calibrated using data with different collision centrality ranges and measurement types. Distributions are normalized to unit integral. Left: Calibrated separately for central (0-10\%, red) and semi-central (20-50\%, grey) \aaa\ collisions. Also shown is the Combined calibration incorporating all centralities (Fig.~\ref{fig:QHatJetHadron}). Middle: For central collisions, additionally restricted to jet--only (blue) or hadron--only (orange) \RAA. The posterior distribution for central collisions in the left panel (red) is shown for comparison. Right: For semi-central collisions, additionally restricted to jet--only (blue) or hadron--only (orange) \RAA. The posterior distribution for semi-central collisions in the left panel (grey) is shown for comparison.}
\label{fig:QHatCentrality}
\end{figure*}

Model calculations show that the space-time temperature distribution of the QGP fireball generated in nuclear collisions depends on collision centrality: the fireball generated in central collisions is initially hotter and lives longer than that in semi-central collisions. This difference in temperature profile will result in quantitatively different jet quenching effects, which may be reflected in the extracted distribution of \qhatTcubed.

Figure~\ref{fig:QHatCentrality} explores this dependence, showing \qhatTcubed\ posterior distributions calibrated separately with data from central (0-10\%) and semi-central (20-50\%) \aaa\ collisions. The central collision posterior distribution for combined data has its largest weight at relatively low values of \qhat, with a long tail extending to larger values.
The semi--central collision posterior distribution for combined data is symmetric and largely overlaps with the distribution from the Combined calibration that incorporates data in the full range of centralities, though with slightly higher mean.

Fig.~\ref{fig:QHatCentrality} shows that the \qhatTcubed\ posterior distributions, when calibrated separately on jet--only or hadron--only \RAA\ data, are each qualitatively similar for central and semi-central collisions. Ref.~\cite{Xie:2022ght} likewise reports \qhatTcubed\ posterior distributions that are consistent for calibrations using inclusive hadron \RAA\ for different centralities, separately for RHIC and LHC data. The difference in the central and semi-central Combined calibrations seen in Fig.~\ref{fig:QHatCentrality} therefore arises predominantly from different relative weights of jet and hadron data. While Fig.~\ref{fig:ExamplePosterior} presents only a limited subset of the data used in this analysis, the selection is broadly representative, showing, for instance, differences in measurement kinematic reach and precision in central and semi--central collisions. These differences must contribute to the close correspondence in Fig.~\ref{fig:QHatCentrality}, left panel, of Combined calibration including both centralities and the semi-central calibration.

The approximate centrality--independence of the \qhatTcubed\ posterior distribution calibrated using jet--only or hadron--only \RAA\ data is itself notable, in light of the differences expected in the the space-time temperature distribution of the QGP fireball for different centralities. This invariance indicates that the difference observed between the jet--only and hadron--only posterior distributions arises predominantly from the different kinematic ranges probed by jets and hadrons, rather than differences in the temperature of the QGP being probed.

Rephrasing this observation in the context of the question posed at the beginning of this section, it shows that the \qhatTcubed\ posterior distribution is indeed not consistent under variation in choice of observable and phase space coverage. Its systematic dependence indicates that this inconsistency arises primarily from sensitivity to the kinematic (\pT) coverage of the probe, and not the modeling of QGP dynamics. This in turn focuses attention on the HTL formulation of \qhatTcubed\ in Eq.~\ref{eq:HTL-q-hat} which, as noted in Sect.~\ref{sec:physicsModel}, is a leading-order approximation, but whose series expansion has been found to have poor convergence properties~\cite{Caron-Huot:2008zna}. Further exploration of this issue, for instance by incorporating higher-order corrections to Eq.~\ref{eq:HTL-q-hat}, is however beyond the scope of this study and will be the focus of future work. 

\subsection{Comparison to previous \qhat\ calibrations}
\label{sect:Compareqhatold}

As noted in Sect.~\ref{sect:Intro}, different determinations of \qhatTcubed, which are based on different theoretical formulations and different choices of inclusive hadron and jet data, may generate constraints that are not directly comparable~\cite{Apolinario:2022vzg}. Their comparison  therefore requires additional analysis. In this section, we focus on comparison of the current results to the previous calibrations of \qhatTcubed\ by the JETSCAPE~\cite{JETSCAPE:2021ehl} and JET~\cite{JET:2013cls} collaborations.

\begin{figure}[tbhp!]
\begin{center}
\includegraphics[width = 0.49\textwidth]{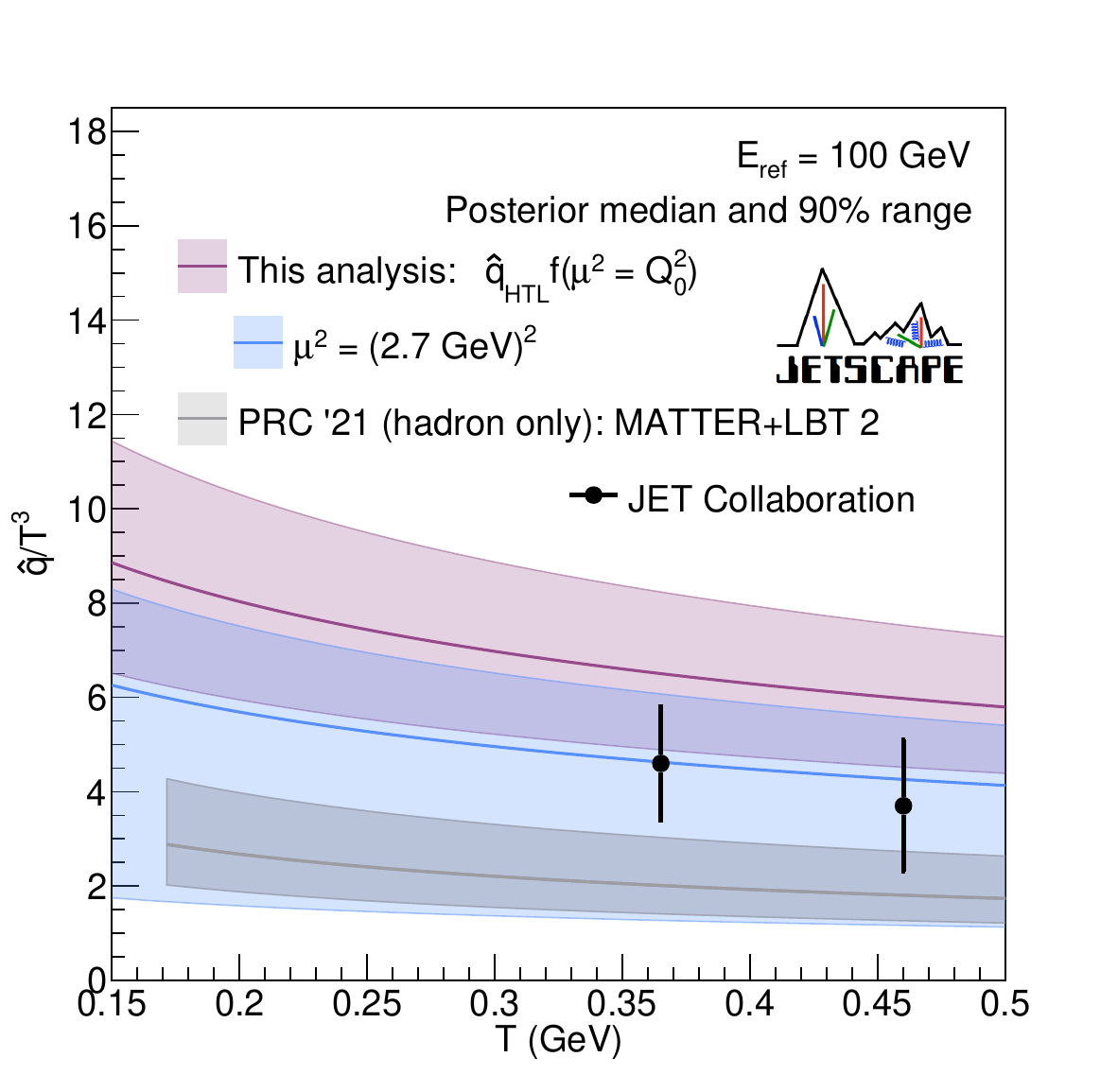}
\end{center}
\caption{Posterior distribution of \qhatTcubed\ as a function of $T$ for the current Combined analysis with $\mu^2=\qswitch^2\sim (1\ \mathrm{GeV})^2$ from Fig.~\ref{fig:QHat}, showing median and 90\% CI limits (purple line and shaded band). Also shown are the  results from \JET~\cite{JET:2013cls} (black data points) and \JETSCAPE~\cite{JETSCAPE:2021ehl} (\MATTER+\LBTtwo\ model of \qhat, Eq.~\eqref{eq:MatterLBTtwo}, grey line and shaded band), which are based solely on inclusive hadron \RAA\ data. The result of the current analysis (Eq.~\eqref{Eq:fmu}) is shown for  $\mu^2=(2.7\ \mathrm{GeV})^2$ (blue line and shaded band), which is the MAP value for the \MATTER+\LBTtwo\ model in Ref.~\cite{JETSCAPE:2021ehl}. See text for details.
}
\label{fig:QHatComparisonQ2}
\end{figure}

Equation~\eqref{Eq:fmu} is the functional form of \qhat in the current analysis, which includes coherence effects that reduce the effective value of \qhat for $\mu \geq \qswitch$. The distribution of \qhatTcubed shown in Fig.~\ref{fig:QHat} from the current Combined analysis is reported for $\mu^2=\qswitch^2$, the medium-induced switching scale, whose distribution in magnitude corresponds to the posterior distribution of the Bayesian analysis; a typical value is $\qswitch^2\sim (1\ \mathrm{GeV})^2$. Fig.~\ref{fig:QHatComparisonQ2} shows the \qhatTcubed distribution compared to the posterior distribution from the previous JETSCAPE Bayesian calibration~\cite{JETSCAPE:2021ehl}, and to values determined by the JET Collaboration~\cite{JET:2013cls}, both of which are based solely on a limited selection of inclusive hadron \RAA data.

The models employed by JET do not incorporate multi--stage energy-loss, scale dependence, or coherence effects. Nevertheless, consistency is observed between JET and the current analysis, taking into account their respective uncertainties.

In the formulation utilized by the previous JETSCAPE calibration (Eq.~\eqref{eq:MatterLBTtwo}), the value of \qhatTcubed\ increases with increasing $\mu$ for $\mu \geq Q_0$, in contrast to the reduction in \qhatTcubed\ with increasing $\mu$ for the current analysis (Eq.~\eqref{Eq:fmu}). Consequently, the Maximum A Posterior (MAP) and 90\% CI intervals of \qswitch\ are different in the two calibrations. The MAP value of $\qswitch^2$ for the current analysis is lower than of the previous calibration ($\sim(2.7\ \mathrm{GeV})^2$)~\cite{JETSCAPE:2021ehl}. Fig.~\ref{fig:QHatComparisonQ2} shows the posterior distribution for the previous JETSCAPE calibration for a value of $\mu^2$ slightly smaller than the MAP value of $(2.7\ \mathrm{GeV})^2$.

In order to compare the two calibrations quantitatively, Fig.~\ref{fig:QHatComparisonQ2} also shows the posterior distribution of \qhatTcubed from the current analysis at $\mu^2=(2.7\ \mathrm{GeV})^2$, using Eq.~\eqref{Eq:fmu}. The 90\% CI interval is wider in this case (blue band) than for $\mu^2=\qswitch^2$ for this analysis (purple band), due to the broad distributions of the calibration parameters $c_1$, $c_2$, and $c_3$, which do not contribute at $\mu^2=\qswitch^2$.
The figure shows that the two calibrations generate consistent posterior distributions, within their respective uncertainties.

It is notable that the two Bayesian calibrations of \qhatTcubed shown in Fig.~\ref{fig:QHatComparisonQ2} are  consistent within uncertainties, despite their markedly different theoretical formulations and the different experimental datasets they employ. This observation raises the question of how to discriminate them based on more detailed analysis. Qualitative visual assessment shows a similar level of agreement between hadron \RAA\ data and the posterior predictive distributions in Ref.~\cite{JETSCAPE:2021ehl} and in this analysis (Fig.~\ref{fig:ExamplePosterior} and Sect.~\ref{app:FullPosteriorPredict}), with good agreement found over significant phase space, but tension found in some regions. However, further exploration of these different \qhat\ formulations in Ref.~\cite{JETSCAPE:2022jer} shows that inclusion of a virtuality--dependent interaction (Eq.~\ref{Eq:fmu}) provides significantly better agreement with the limited dataset used in Ref.~\cite{JETSCAPE:2021ehl}, and with the broader dataset used in this analysis. Additional Bayesian Inference tools to discriminate different models quantitatively~\cite{Phillips:2020dmw,JETSCAPE:2020mzn} will be explored in future calibrations of \qhat.

%% file: Summary.tex
\section{Summary}
\label{sect:Summary}

The \JETSCAPE\ collaboration reports a new, multi-observable determination of the jet transport coefficient \qhat, using all available inclusive hadron and jet suppression data from RHIC and the LHC. The model of the QGP bulk medium and its evolution is based on parameters determined by a previous Bayesian calibration of soft-sector observables. Virtuality-dependent jet quenching is implemented in a multi-stage model.

The combined calibration of \qhatTcubed, using both inclusive hadron and jet \RAA\ data, describes the data well over a significant phase space, though with tension in some regions. The posterior distribution of \qhatTcubed\ increases with decreasing $T$, consistent with some other determinations of \qhat\ from inclusive hadron and jet data. 

Additional differential studies explore the interplay of hadron and jet \RAA\ data in constraining the posterior distributions. High-\pT\ hadron data (roughly, $\pT>30$~\gev) provide consistent posterior constraints as the jet \RAA\ data, much of which covers the range $p_\mathrm{T,jet}>50$~\gev; these observables evidently probe jet quenching in similar partonic phase space. However, the posterior distribution from calibration with lower-\pT\ hadron \RAA\ is not consistent, indicating that the model dependence of \qhatTcubed\ on parton energy is not fully accurate. 

The centrality dependence of the posterior distributions, and their further classification based on jet--only or hadron--only \RAA\ measurements, likewise indicates that the most significant source of tension in the comparison of the current model to data is the functional dependence on parton energy $E$. Improving this model description, and exploring alternative modeling approaches, is likewise the subject of future work.

The calibration in this analysis is consistent with that of the previous JETSCAPE calibration of \qhatTcubed, which is based on a formulation of \qhat\ with different functional dependence on parton virtuality and using data corresponding to a subset of the hadron \RAA\ data in this analysis. The consistency is manifest when the \qhatTcubed{} is evolved in virtuality $\mu$ to the scale $Q_0 \approx 2.7$~GeV of the prior analysis [using Eq.~\eqref{Eq:fmu}].

The analysis presented represents a significant step towards the long-term goal of a comprehensive multi-observable Bayesian calibration of jet quenching data to constrain fundamental transport properties of the Quark-Gluon Plasma. However, as noted above, it has raised several important questions that require resolution with future work in order to achieve this goal. An equally important, long--standing issue in the field is the specification of meaningful theoretical and modeling uncertainties, to be used in the likelihood calculations that are at the heart of Bayesian Inference. While this issue is likewise beyond the scope of the present work, the analysis presented in this paper serves to highlight the urgent need for progress in this area as well.

%% file: Appendix.tex
\appendix

\section{Full parameter posterior distributions}
\label{app:FullParaPosteriors}

\begin{figure}[tbhp!]
\begin{center}
\includegraphics[width = 0.45\textwidth]{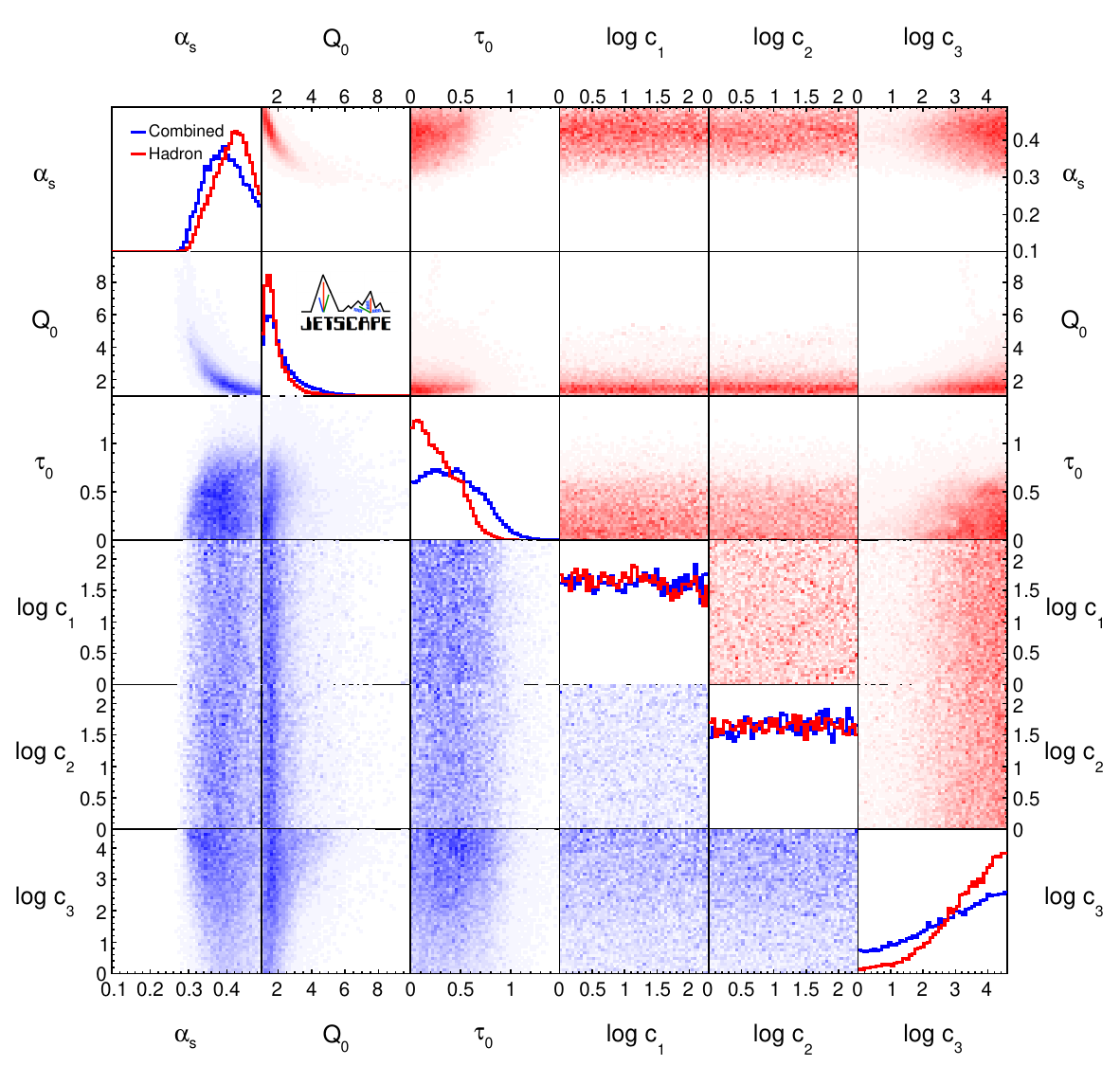}
\end{center}
\caption{Same as Fig.~\ref{fig:PosteriorParameter} but also showing the posterior distributions and correlations for parameters $c_1$ and $c_2$.}
\label{fig:PosteriorParameterAlt}
\end{figure}

Figure~\ref{fig:PosteriorParameterAlt} shows the same parameter posterior distributions and correlations as Fig.~\ref{fig:PosteriorParameter}, in addition including those for parameters $c_1$ and $c_2$.

\section{Complete set of posterior predictive distributions}
\label{app:FullPosteriorPredict}

\begin{figure*}
    \centering
\includegraphics[height=0.09\textheight]{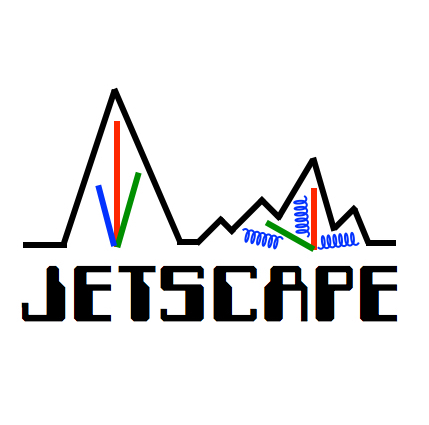}
\includegraphics[height=0.09\textheight]{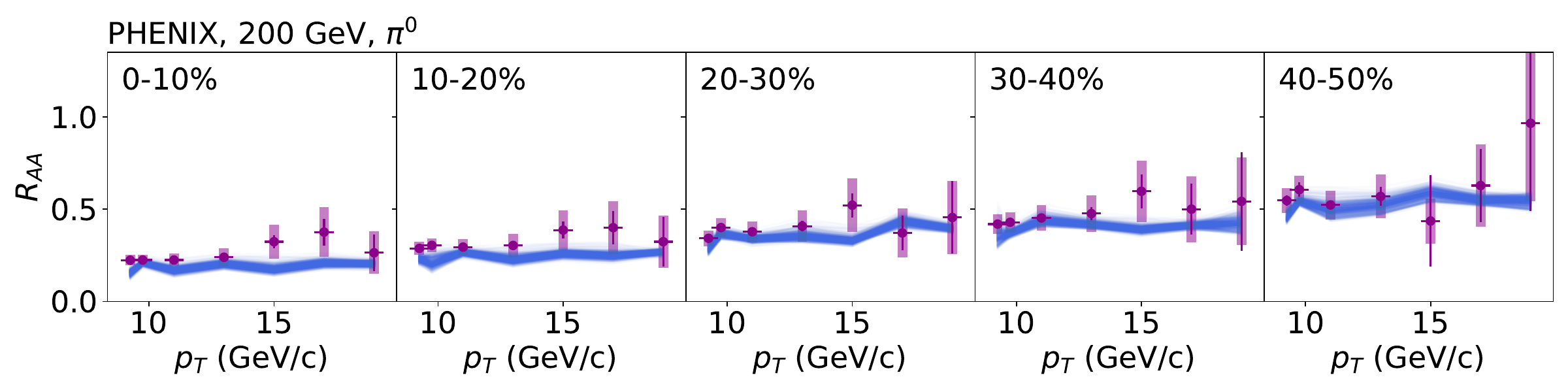}
\includegraphics[height=0.09\textheight]{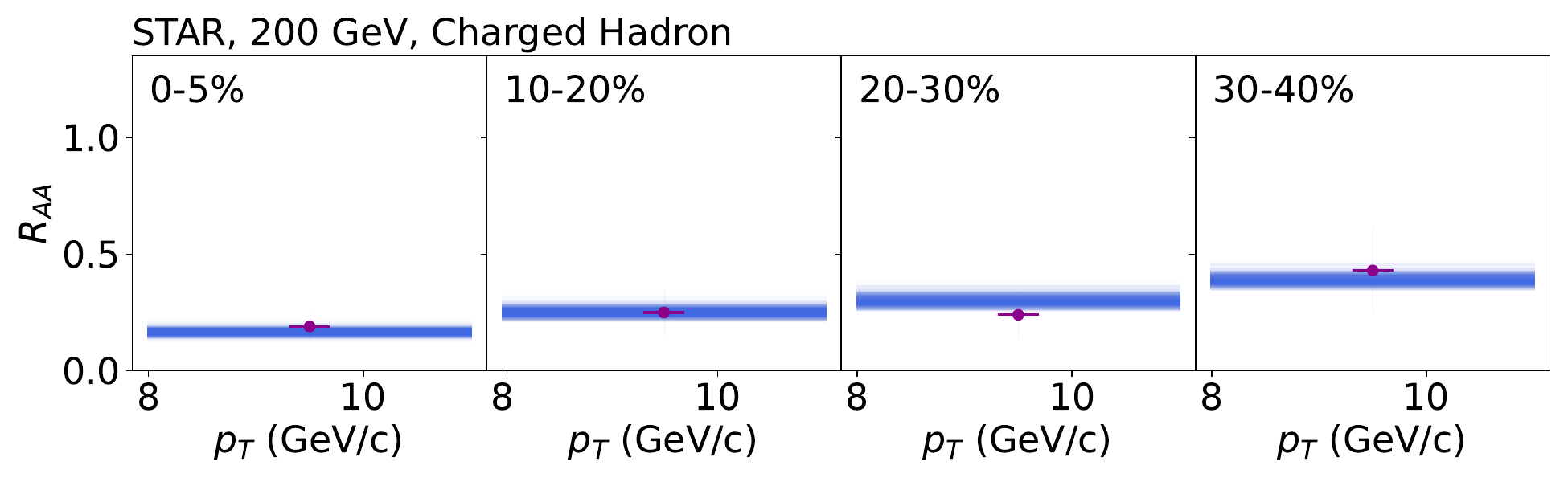}
\includegraphics[height=0.09\textheight]{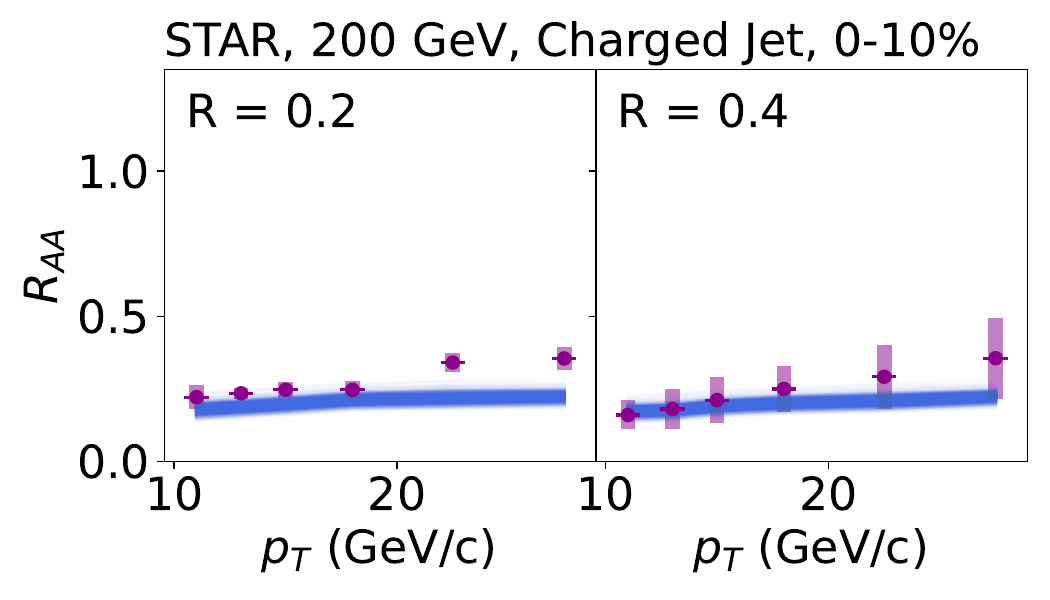}
\includegraphics[height=0.09\textheight]{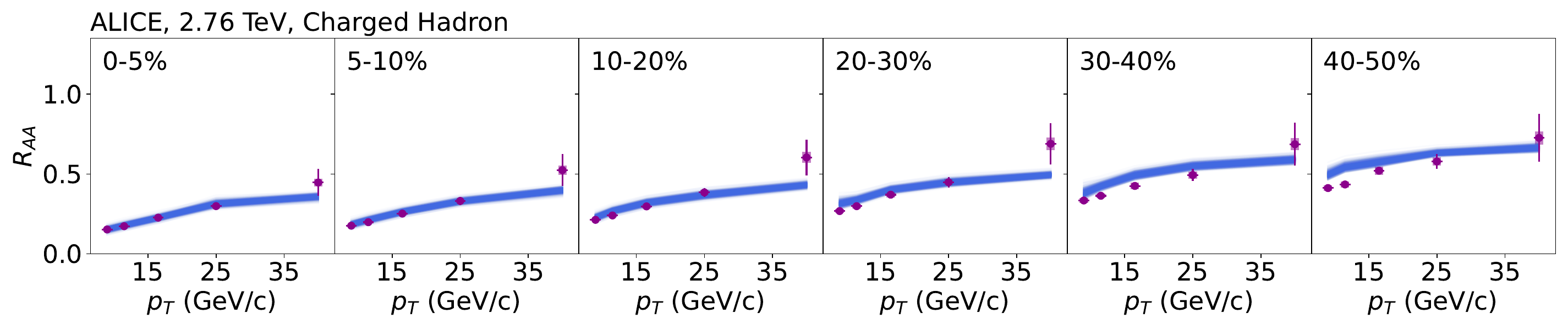}
\includegraphics[height=0.09\textheight]{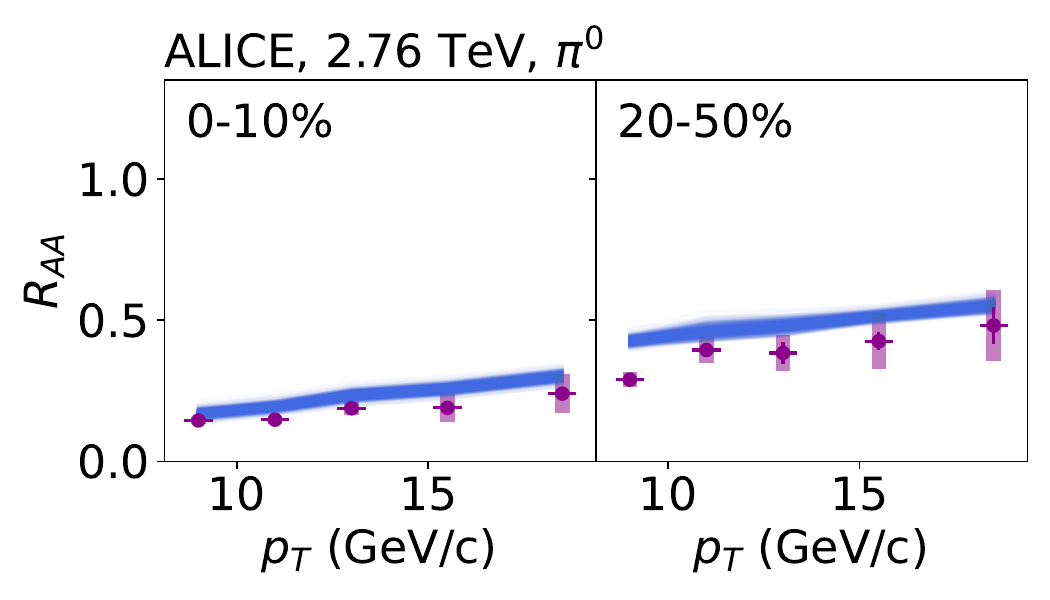}
\includegraphics[height=0.09\textheight]{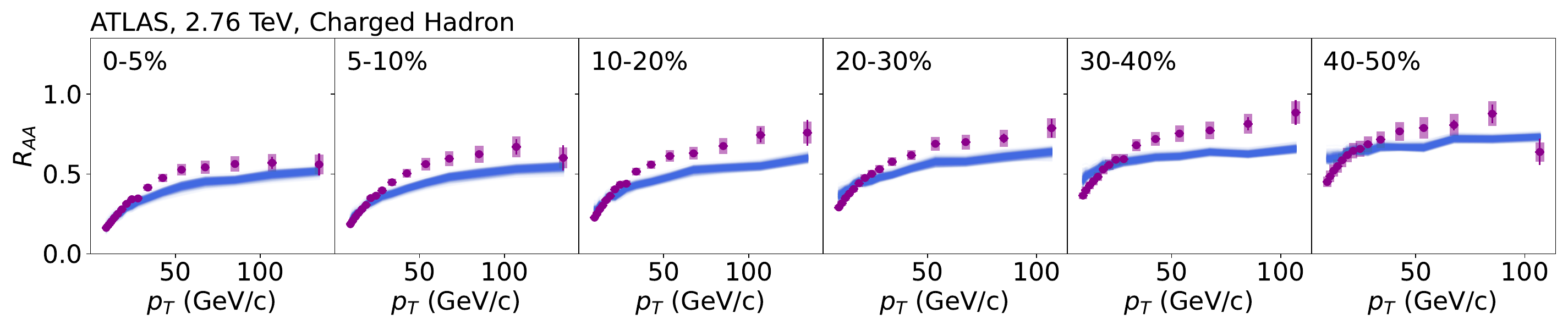}
\includegraphics[height=0.09\textheight]{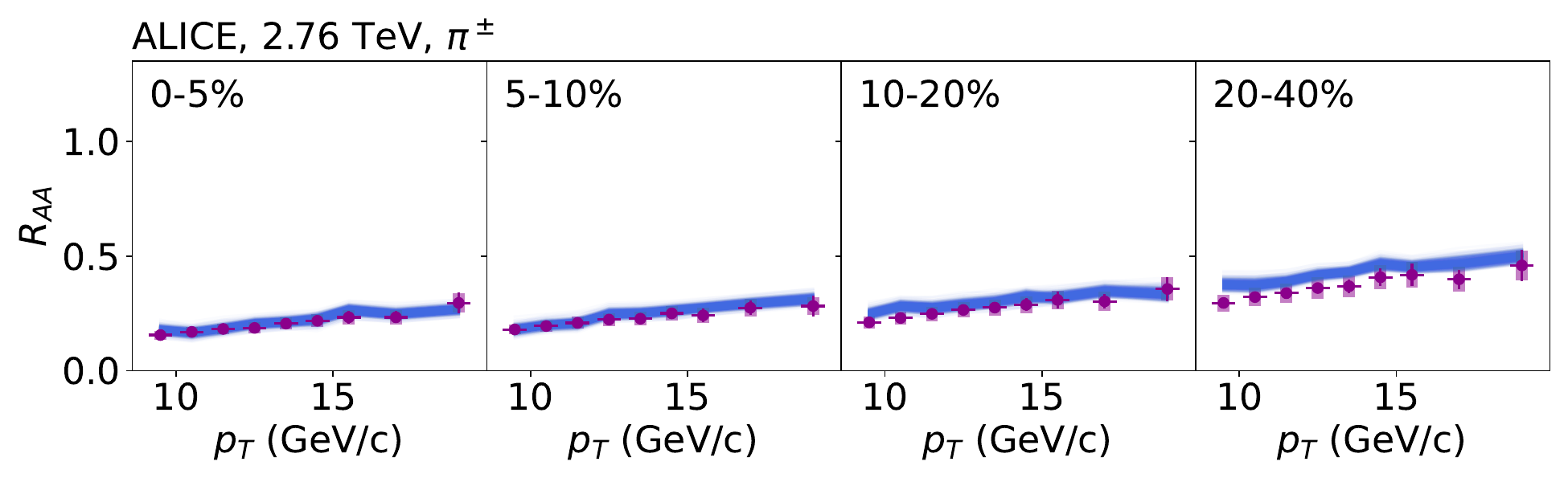}
\includegraphics[height=0.09\textheight]{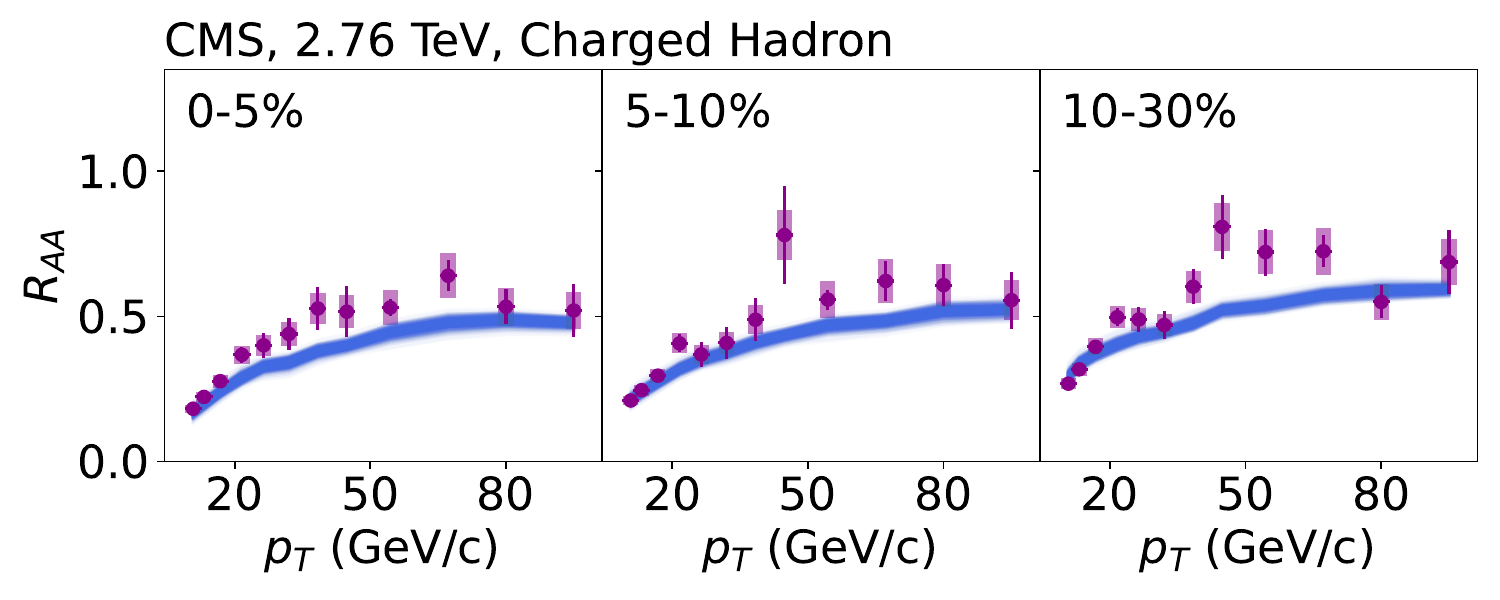}
\includegraphics[height=0.09\textheight]{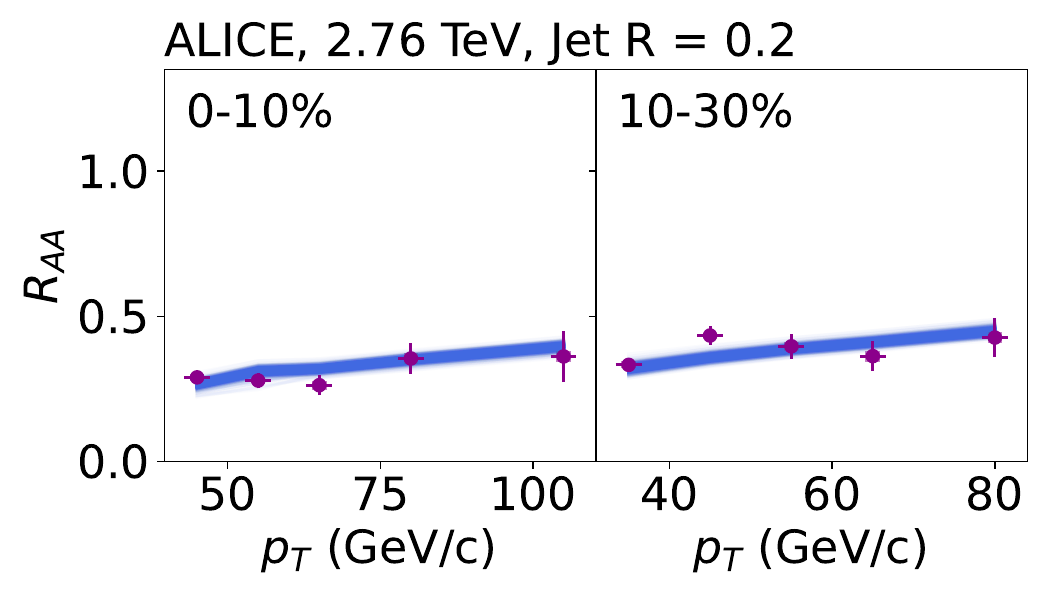}
\includegraphics[height=0.09\textheight]{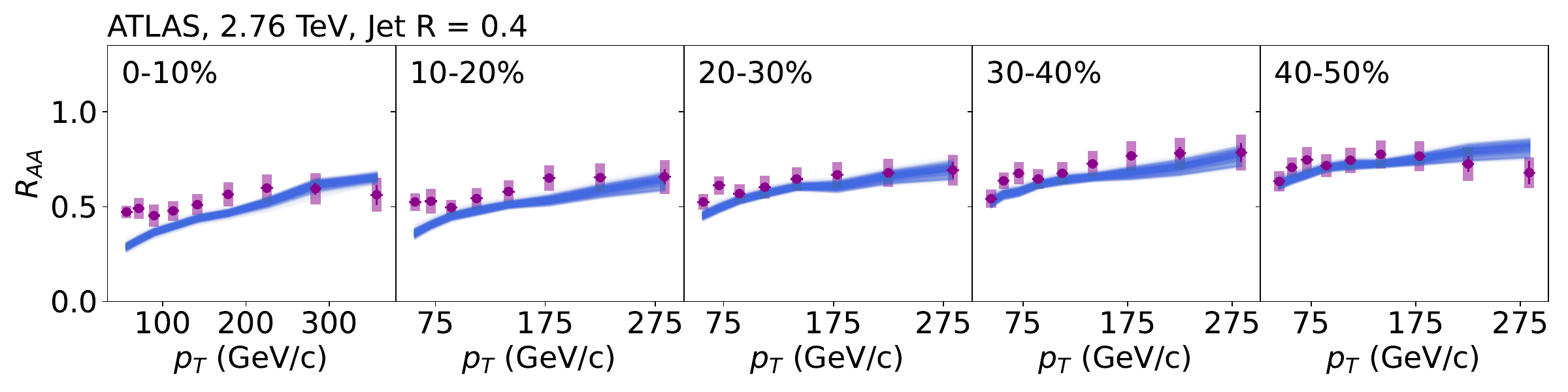}
\includegraphics[height=0.09\textheight]{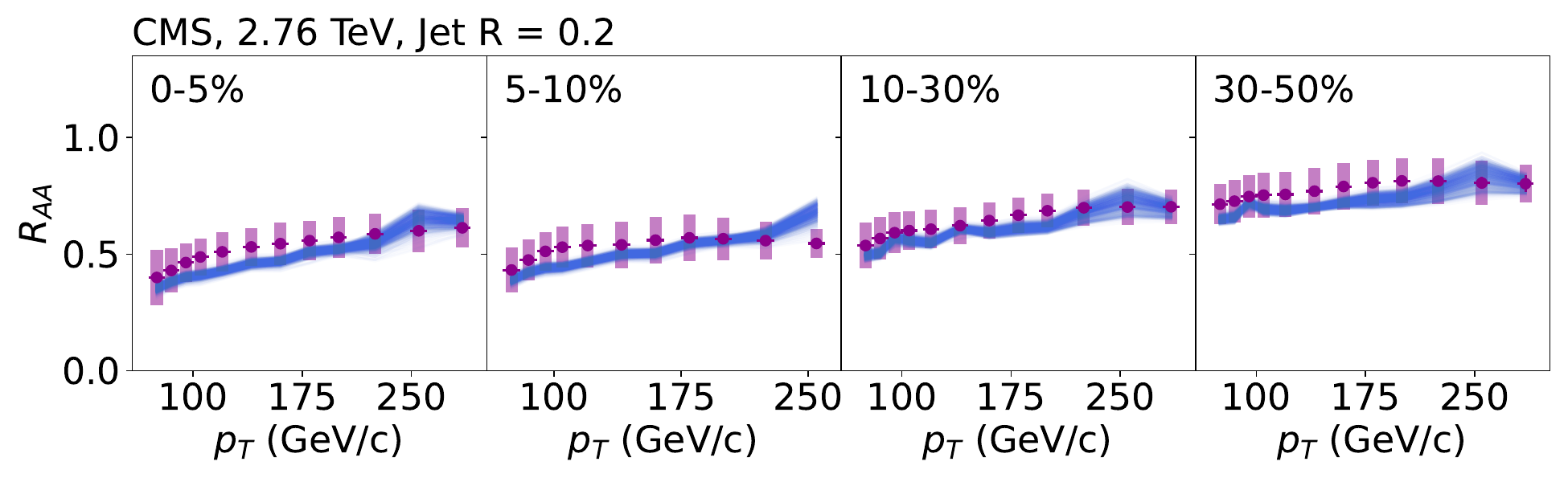}
\includegraphics[height=0.09\textheight]{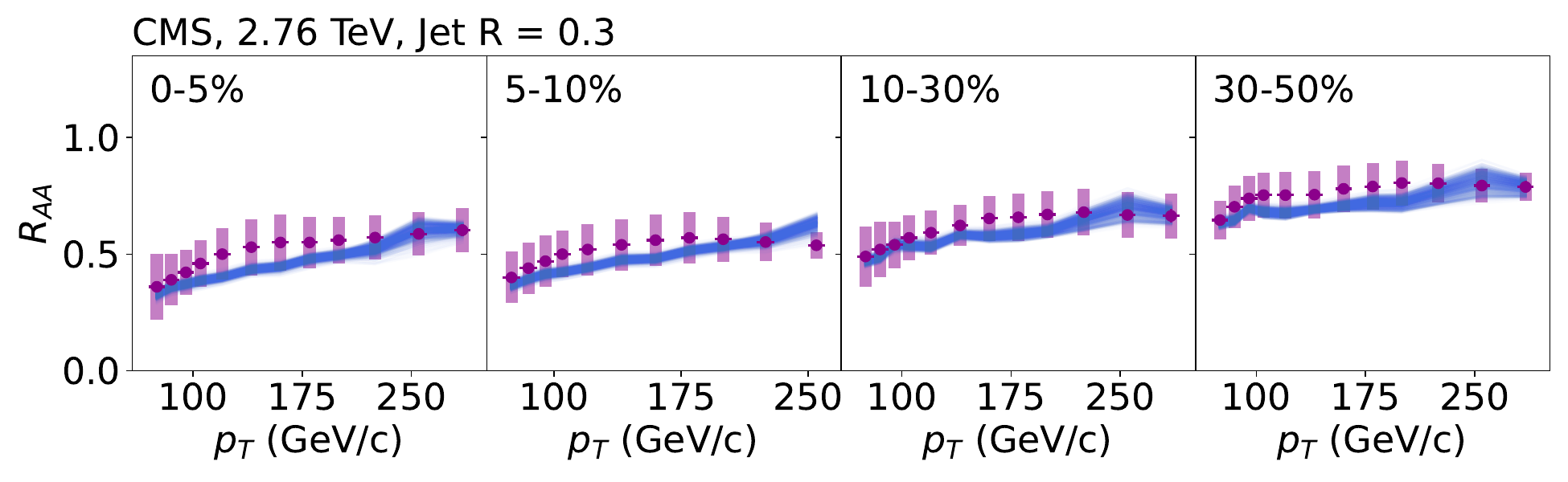}
\includegraphics[height=0.09\textheight]{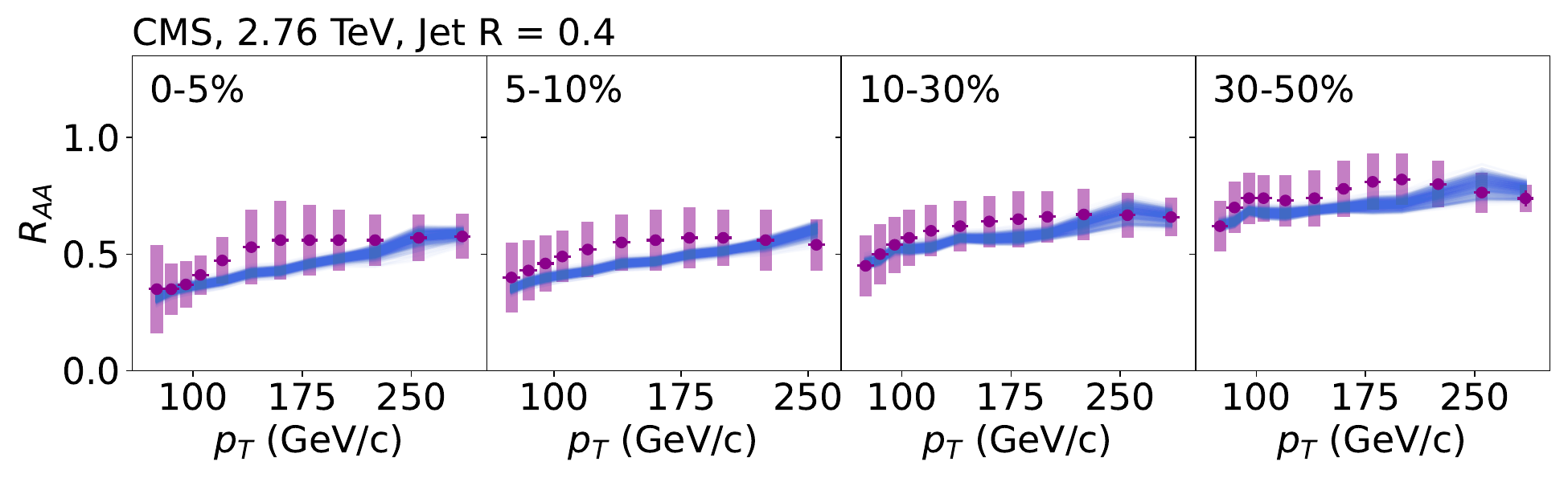}
\includegraphics[height=0.09\textheight]{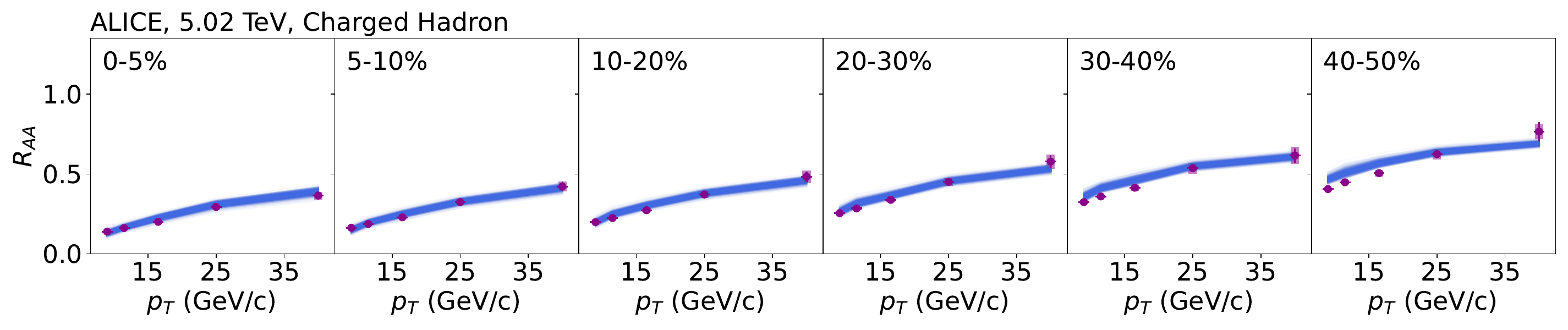}
\includegraphics[height=0.09\textheight]{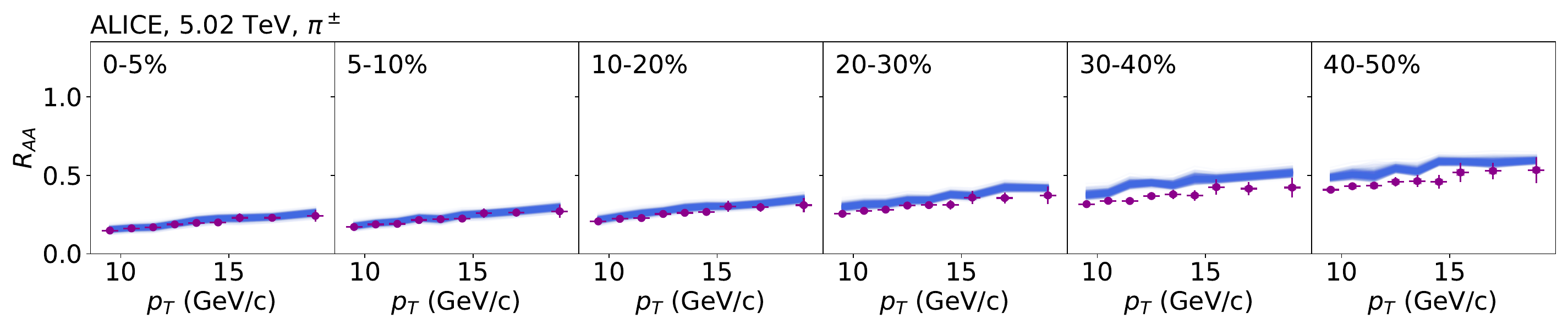}
\includegraphics[height=0.09\textheight]{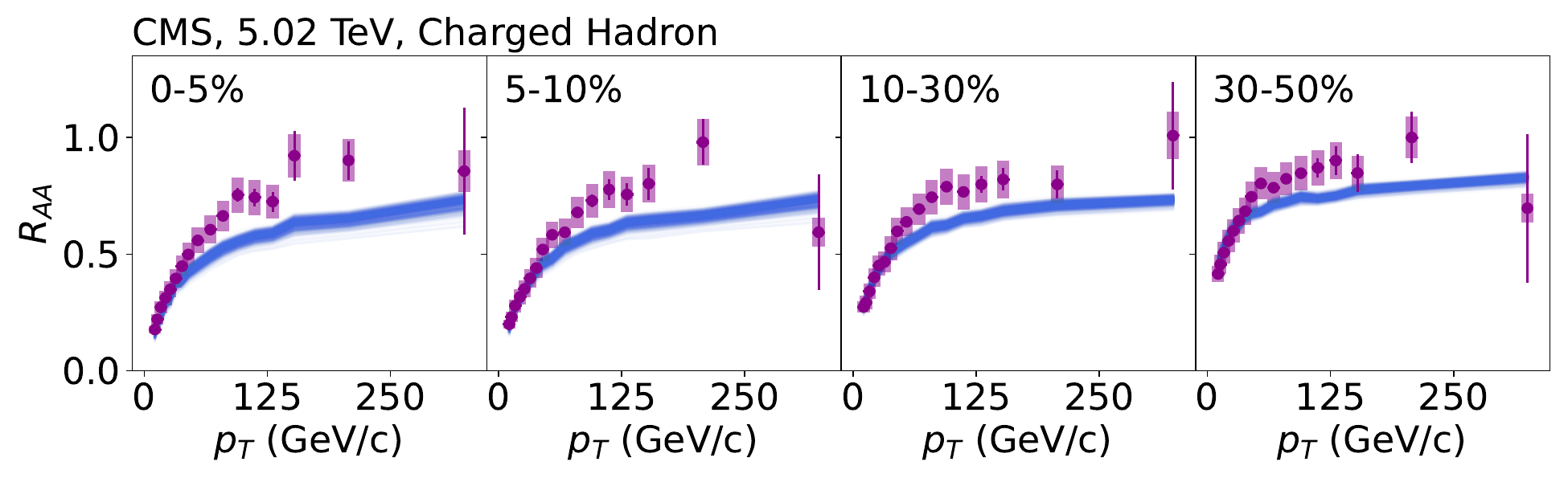}
\includegraphics[height=0.09\textheight]{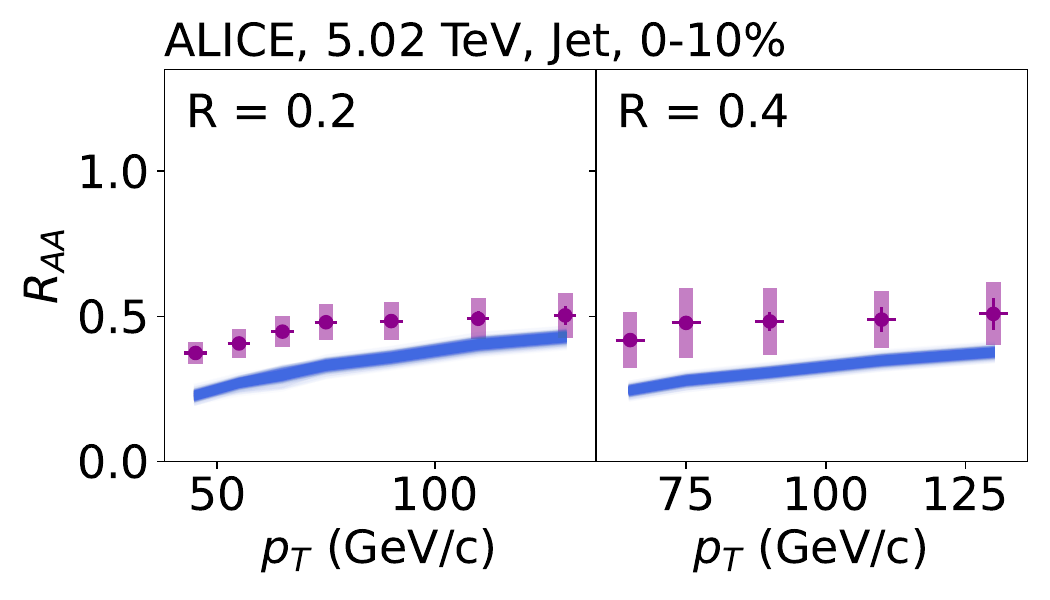}
\includegraphics[height=0.09\textheight]{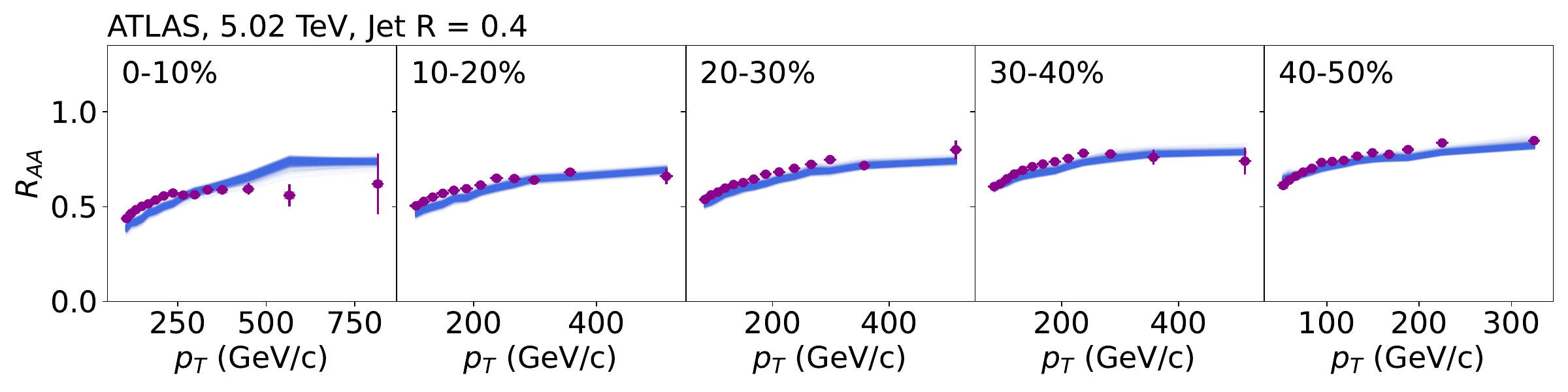}\\
\includegraphics[height=0.09\textheight]{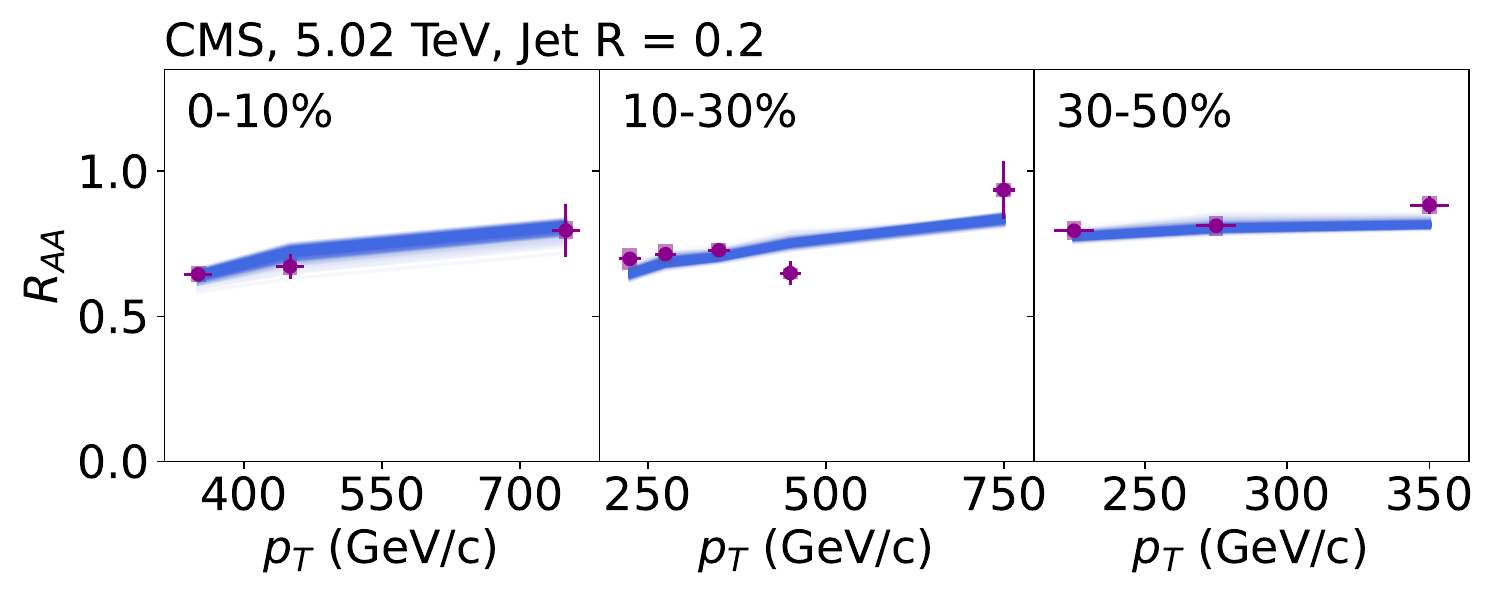}
\includegraphics[height=0.09\textheight]{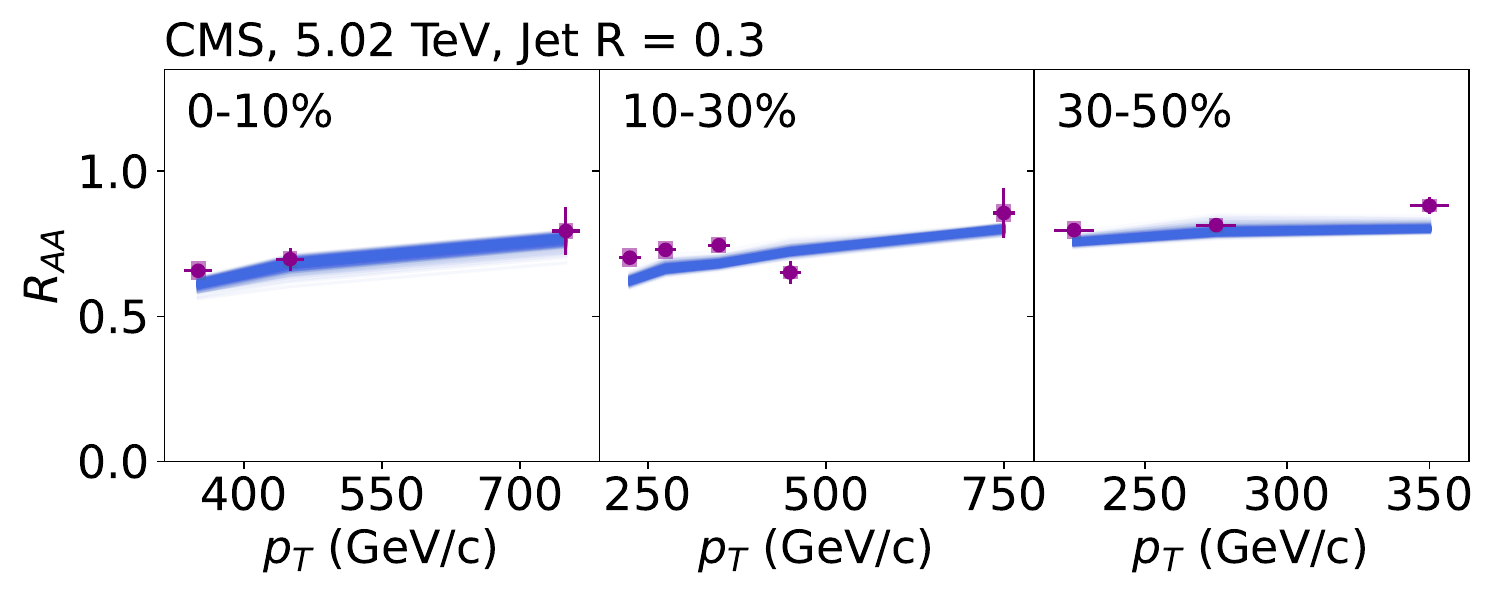}
\includegraphics[height=0.09\textheight]{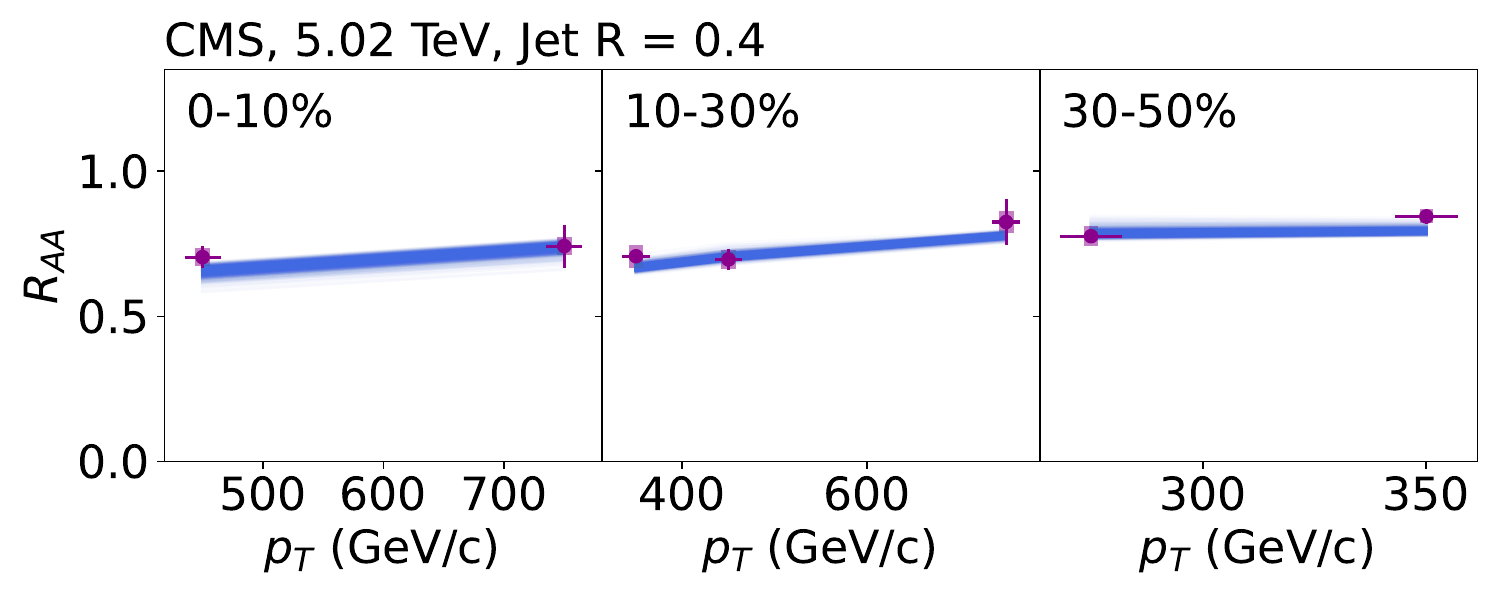}
\includegraphics[height=0.09\textheight]{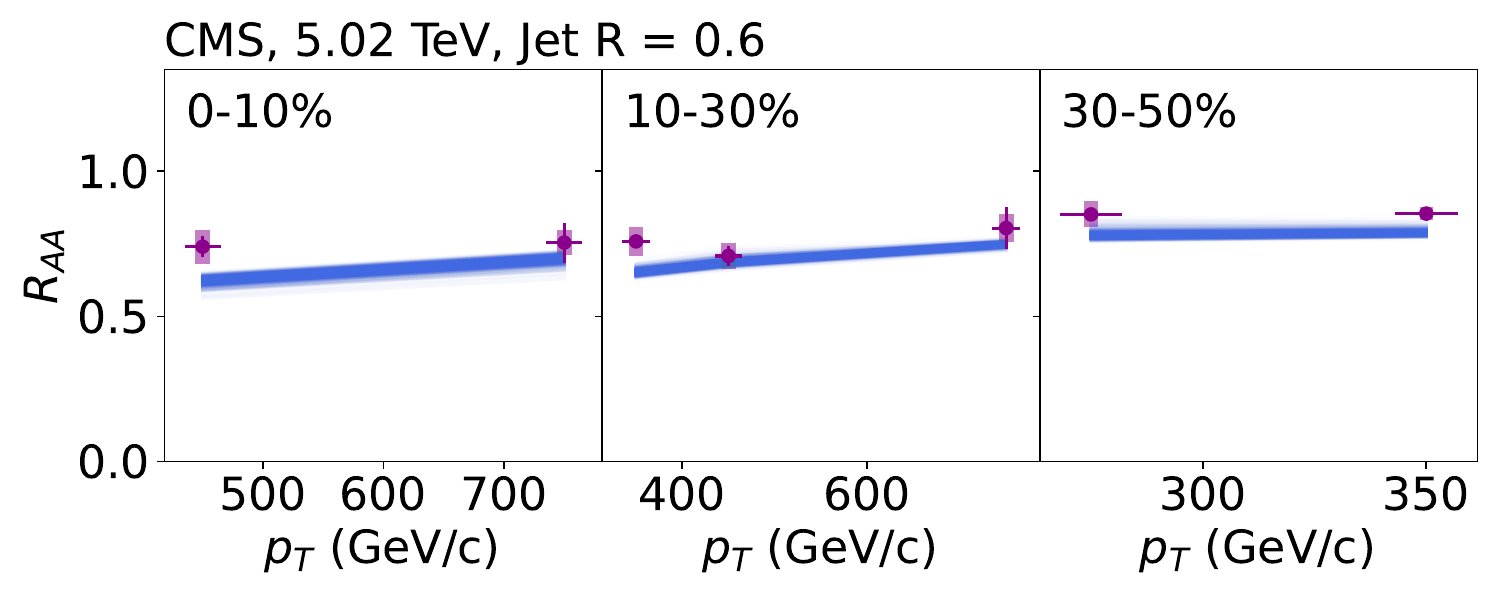}
\includegraphics[height=0.09\textheight]{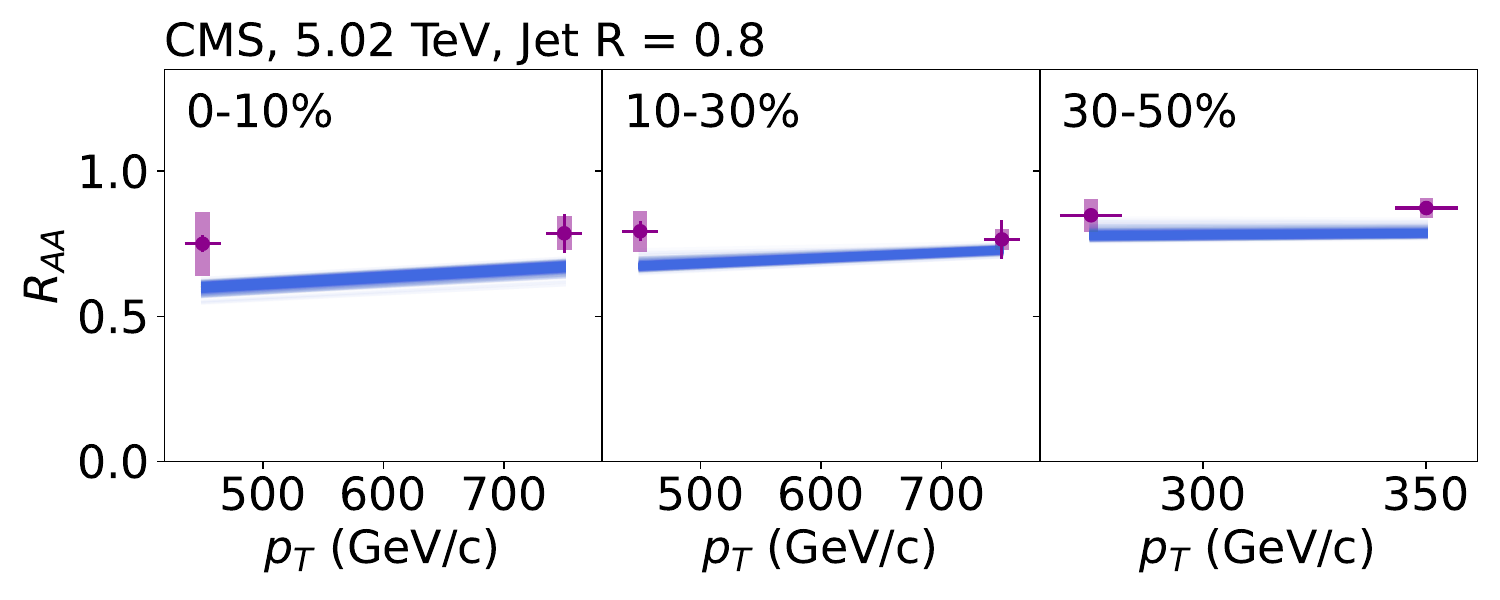}
\includegraphics[height=0.09\textheight]{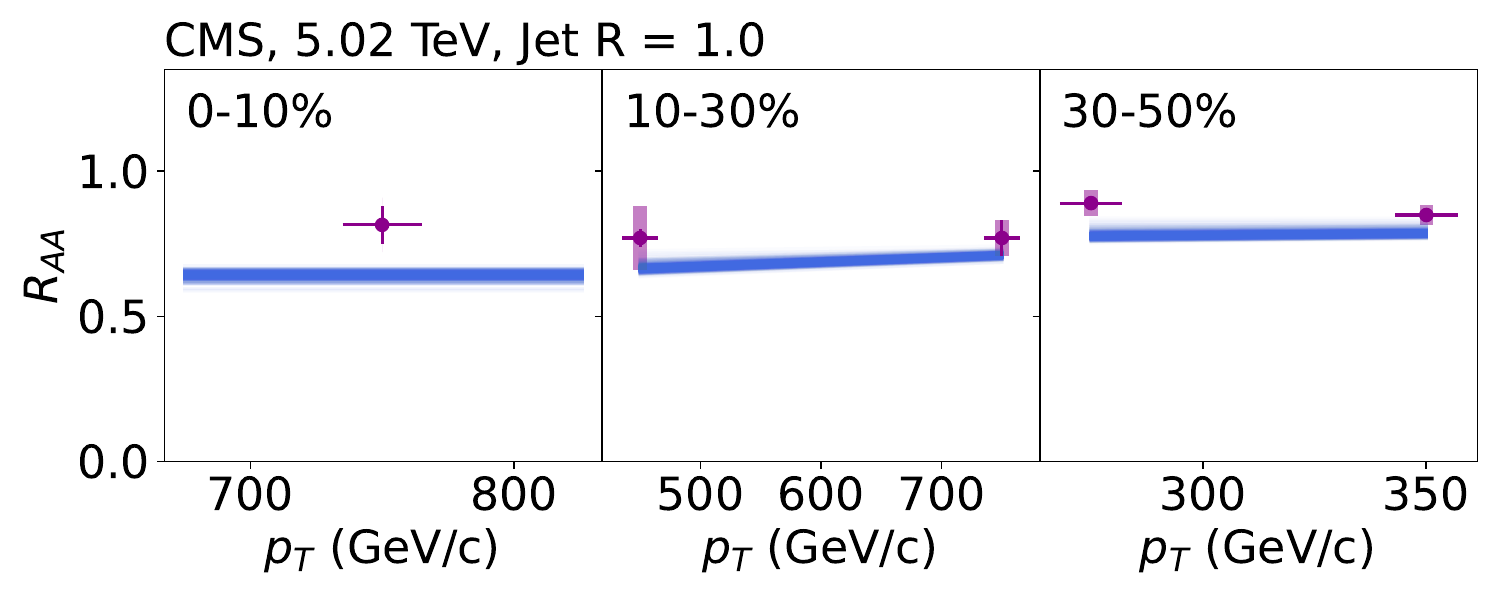}
\caption{All measurements of hadron and jet \RAA at $\sqrtsNN= 0.2$, 2.76, and 5.02 TeV at various centralities (purple) considered in this analysis, compared to posterior predictive distributions for the Combined analysis (blue).}
    \label{fig:FullPosterior}
\end{figure*}

Figure~\ref{fig:ExamplePosterior} shows a representative selection of jet and \RAA\ data compared to the posterior predictive distributions. The complete set of jet and \RAA\ data used in the Combined analysis, along with the posterior predictive distributions, is shown in Fig.~\ref{fig:FullPosterior} for comparison.

